\documentclass[aps,superscriptaddress,prl,reprint]{revtex4-1}

\label{pre:packages}
\usepackage{siunitx}
\usepackage[normalem]{ulem}
\usepackage{graphicx}
\usepackage{amsmath,amssymb}
\usepackage[inline]{enumitem}
\usepackage{array}
\usepackage{xcolor}
\usepackage{comment}

\usepackage{hyperref}

\hypersetup{
  colorlinks=true,
  citecolor=blue,
  linkcolor=blue,
  urlcolor=blue,
}

\usepackage[capitalize]{cleveref}

\newif\ifsol
\label{pre:ifsol}
\solfalse

\usepackage{tabularx}
\newcolumntype{Y}{>{\centering\arraybackslash}X}

\ifsol

\usepackage[prefix=]{xcolor-solarized}
\colorlet{gray}{base01}
\colorlet{hl}{base02}
\makeatletter
\newcommand{\globalcolor}[1]{%
  \color{#1}\global\let\default@color\current@color
}
\makeatother
\AtBeginDocument{
  \globalcolor{base0}
  \pagecolor{base03}
}

\else

\usepackage{xcolor}
\colorlet{hl}{green!50!}
\colorlet{green}{green!50!black!}
\colorlet{blue}{blue!55!black!}

\fi

\label{pre:macros}

\newcommand{\bmb}{BaMn$_2$Bi$_2$}
\newcommand{\bma}{BaMn$_2$As$_2$}
\newcommand{\bmpn}{BaMn$_2${\itshape Pn}$_2$}
\newcommand{\bfa}{BaFe$_2$As$_2$}

\renewcommand{\vec}[1]{\mathbf{#1}}
\DeclareSIUnit[number-unit-product = \,]{\unitcell}{[\mathrm{unit\ cell}]}
\newcommand{\tmin}{T_{\mathrm{min}}}

\newcommand{\EF}{E_{\mathrm{F}}}

\mathchardef\mhyphen="2D

\newcommand{\vecHab}{\vec{H}_{ab}}
\newcommand{\Hab}{H_{ab}}

\newcommand{\vecHa}{\vec{H}_a}
\newcommand{\Ha}{H_{a}}

\newcommand{\vecHc}{\vec{H}_c}
\newcommand{\Hc}{H_{c}}

\begin{document}

\title{Magnetic-field-induced Anderson localization\\
  in orbital selective antiferromagnet \bmb{}}
\label{sec:authorlist}


\author{Takuma Ogasawara}
\affiliation{Department of Physics, Graduate School of Science, Tohoku University, \\
  6-3 Aramaki, Aoba, Miyagi, Japan}

\author{Kim-Khuong Huynh}
\email{huynh.kim.khuong.b4@tohoku.ac.jp}
\affiliation{Advanced Institute for Materials Research (WPI-AIMR), Tohoku University, 1-1-2 Katahira, Aoba, Sendai, Miyagi, Japan}

\author{Stephane Yu Matsushita}
\affiliation{Advanced Institute for Materials Research (WPI-AIMR), Tohoku University, 1-1-2 Katahira, Aoba, Sendai, Miyagi, Japan}

\author{Motoi Kimata}
\affiliation{Institute for Materials Research, 1-1-2 Katahira, Aoba, Sendai, Miyagi, Japan}

\author{Time Tahara}
\affiliation{Center for Advanced High Magnetic Field Science, Graduate School Science, Osaka University,
  1–1 Machikaneyama, Toyonaka, Osaka, Japan}

\author{Takanori Kida}
\affiliation{Center for Advanced High Magnetic Field Science, Graduate School Science, Osaka University,
  1–1 Machikaneyama, Toyonaka, Osaka, Japan}

\author{Masayuki Hagiwara}
\affiliation{Center for Advanced High Magnetic Field Science, Graduate School Science, Osaka University,
  1–1 Machikaneyama, Toyonaka, Osaka, Japan}

\author{Denis Ar\v{c}on}
\affiliation{Faculty of mathematics and physics, University of Ljubljana,
  Jadranska c. 19, 1000 Ljubljana, Slovenia}
\affiliation{Jozef Stefan Institute,
  Jamova c. 39, 1000 Ljubljana, Slovenia}

\author{Katsumi Tanigaki}
\email{katsumi.tanigaki.c3@tohoku.ac.jp}
\affiliation{Advanced Institute for Materials Research (WPI-AIMR), Tohoku University, 1-1-2 Katahira, Aoba, Sendai, Miyagi, Japan}
\affiliation{BAQIS, Bld. 3, No.10 Xibeiwang East Rd., Haidian District, Beijing 100193, China}

\date{December 28, 2017}

\begin{abstract}
  We report a metal-insulator transition (MIT) in the half-filled multiorbital antiferromagnet (AF) \bmb{} that is tunable by magnetic field perpendicular to the AF sublattices.
  Instead of an Anderson-Mott mechanism usually expected in strongly correlated systems, we find by scaling analyses that the MIT is driven by an Anderson localization.
  Electrical and thermoelectrical transport measurements in combination with electronic band calculations reveal a strong orbital-dependent correlation effect, where both weakly and strongly correlated $3d$-derived bands coexist with decoupled charge excitations.
  Weakly correlated holelike carriers in the $d_{xy}$-derived band dominate the transport properties and exhibit the Anderson localization, whereas other $3d$ bands show clear Mott-like behaviors with their spins ordered into AF sublattices.
  The tuning role played by the perpendicular magnetic field supports a strong spin-spin coupling between itinerant holelike carriers and the AF fluctuations, which is in sharp contrast to their weak charge coupling.
\end{abstract}

\maketitle

\paragraph{Introduction}---~%
A metal-insulator transition (MIT) can occur in a clean solid as a consequence of strong electron-electron correlations that open a Mott pseudogap in the excitation spectrum \cite{mott2004metal}.
On the other hand, quantum interferences arising from disorder in a non-interacting system can destroy all extended states and induce an Anderson localization (AL) transition \cite{anderson1958}.
The intertwining of these two mechanisms within the same system generates rich phase diagrams containing coexisting complex phases and intriguing critical behaviors at the phase boundaries \cite{dobrosavljevic2012conductor,byczuk2010}.
Theoretical calculations of correlated single band models at half-filling have found that correlated metallic phases can be enhanced in the regime of weak disorder and correlation \cite{dobrosavljevic2012conductor,byczuk2010}.
Intricate Anderson-Mott insulating phases with suppressed Mott-like antiferromagnetic order (AF) \cite{byczuk2009} and/or coexistence of spatially segregated Mott and Anderson regions have been proposed in the regime of strong disorder and correlation \cite{aguiar2009}.

The participation of multiple orbitals often found in transition metal compounds can make the above competitions even more interesting.
Orbital selective Mottness scenarios have been recently put forward to discuss the cases of coexisting weakly and strongly correlated energetic bands, both of which are derived from the same partially filled $d$-shell \cite{demedici2014,demedici2011}.
Unfortunately, Anderson-Mott and AL transition in the presence of orbitally selective Mottness have been experimentally and theoretically unexplored due to the lack of suitable model systems.

\begin{figure}
  \includegraphics[width = .48\textwidth]{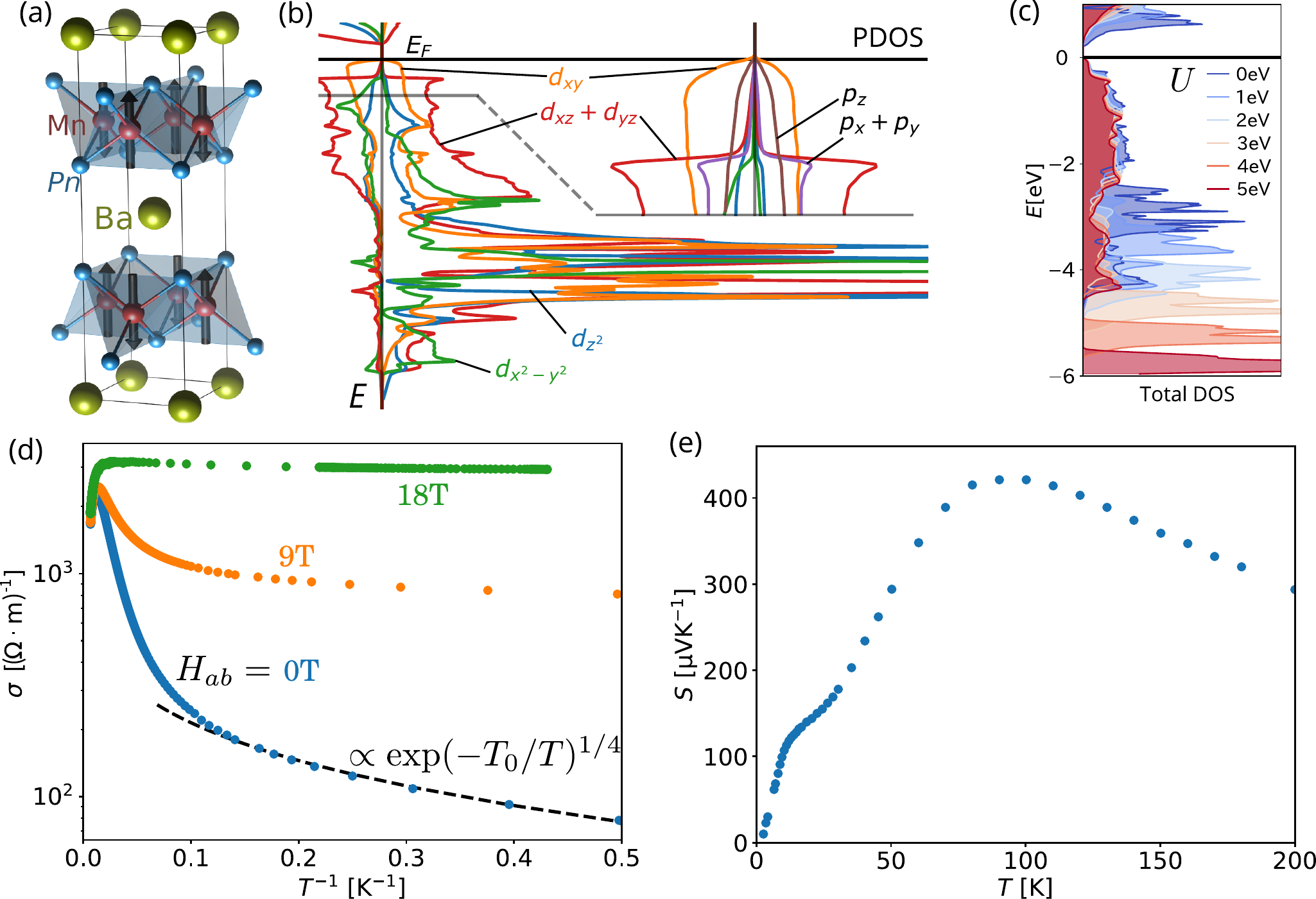}
  \caption{(Color online)
    (a) The crystallographic and magnetic structure of \bmpn{}.
    (b) Spin-polarized PDOSs of Mn's $3d$ and Bi's $6p$ orbitals calculated for $U=0$ at a Mn site in the spin-up AF sublattice.
    The inset at the right hand side enlarges the region near the $\EF$.
    (c) $U$ heavily affects all bands except the $d_{xy}$-derived one close to the $\EF$.
    (d) The electrical conductivity of \bmb{} measured in perpendicular magnetic field $\vecHab \perp \hat{c}$ configuration.
    The black dashed line is the fit to VRH model.
    (e) The temperature dependence of the Seebeck coefficient of \bmb{} at zero magnetic field.
  }
  \label{fig:dos}
\end{figure}

In the present paper, we combine experimental observations and band calculations to show that with the help of orbital degrees of freedom, the dominant role in such intriguing cases can be taken by an AL mechanism even in the presence of strong electronic correlations.
  Our model compound is single crystalline \bmb{} [Fig.~\ref{fig:dos}(a)], the end-member in the family of half-filled G-type AF \bmpn{}'s ({\itshape Pn} stands for As, Sb, and Bi) \cite{huynh2019,ogasawara2021,jansa2021}.
  In these materials, structurally similar to their sister \bfa{} with a $3d^6$ electronic shell \cite{demedici2014}, five $3d$ orbitals are essential to the electronic properties.
  The correlation strength varies greatly depending on the Mn's $3d$ orbitals, even though \bmpn{} is at half-filling \cite{zingl2016,craco2018}.
    
  \bmb{} exhibits a clear orbital-dependence, as its $d_{xy}$-derived band is virtually unaffected by electronic correlation whereas the other orbitals show very strong Mott-insulating behavior with AF order [Fig.~\ref{fig:dos}(b)-(c)].
  The holelike charges in the $d_{xy}$-derived band exhibit a bad metallic state at high temperatures, but with cooling down undergo a localization transition that renders a weak insulator.
  Intriguingly, MIT is tunable by a magnetic field that is \emph{perpendicular} to the G-AF axis ($\vecHab$) so that the material becomes metallic at high magnetic field strength $\Hab$.
  By employing scaling analyses for the conductivities at different $\Hab$'s, we found that the critical exponent of MIT is $\nu \approx 1.4$.
  This value corresponds to an AL transition, but not a Anderson-Mott transition, in the unitary universal class and therefore reaffirms the absence of correlation effect in the $d_{xy}$-derived band.
  On the other hand, the unusual tuning role played by $\vecHab$ supports a picture of strong coupling between the spin angular momenta of $d_{xy}$ holes and the AF fluctuations.
  The contrast in coupling strengths, weak for charge and strong for spin excitations, is a reminiscence of the scenario of spin-charge separation recently proposed for multiobital Hund's metal \cite{demedici2014,deng2019}.
  We also describe a hidden spin-glass-like state that may associate with the magnetism arising from the $d_{xy}$ holes in the AL regime.

\paragraph{Multiorbital electronic structure and magnetism}---~%
\label{sec:orbital-energy}
Fig.~\ref{fig:dos}(b) shows the band structure of \bmb{} as obtained from our spin-polarized band calculations \footnote{See supplementary information} by using program code WIEN2k \cite{blaha_wien2k_2020}.
Both ``core'' and itinerant electronic states coexist.
The ``core'' states locate far away from $\EF$, and their spins are well-ordered in to G-AF sublattices.
These AF-polarized bands show clear Mott-like behaviors because they are effectively pushed down by increasing the correlation strength $U$ [Fig. \ref{fig:dos}(c)].

Much more important for  transport properties is the band at the vicinity of $\EF$.
The partial density of states (PDOS) [Fig.~1(b)] show that the contribution from the $3d_{xy}$ orbital dominates this band, being $60$ and $10$ times larger than those from the Mn's $d_{xz/yz}$ and Bi's $p_{z}$ orbitals, respectively.
This $d_{xy}$-derived band resembles that of a paramagnetic metal.
Its wide band width that favors itinerant holes and the PDOSs of ``spin-up'' and ``spin-down'' states are almost equivalent.
Interestingly, we found that the $d_{xy}$-derived band remains virtually unaffected even for extremely large $U$ [Fig. \ref{fig:dos}(c)]. 
Our band calculations show a strong orbital-selective correlation effect in \bmb{} and are in a good agreement with the previous calculations for the sister compound \bma{} \cite{craco2018,zingl2016}.
Whereas other \bmpn{}'s remain insulating under high magnetic fields, \bmb{} is in the vicinity of an intriguing field-tuned MIT.
The $d_{xy}$-derived band at the vicinity of the $\EF$ is the playground for interesting transport and magnetism, as described below.

\paragraph{Localization in transport properties and Delocalization via $\vecHab$}---~%
At high temperatures ($T$) and zero magnetic field, \bmb{} exhibits a bad metal behavior which then changes to insulating at $\tmin \approx \SI{83}{\kelvin}$.
Fig.~\ref{fig:dos}(d) shows that the conductivity $\sigma$ follows an insulating Arrhenius law at $T < \tmin$ and then a 3-dimensional variable range hoping (VRH) $\sigma \propto \exp\left [-(T_0/T)^{1/4}\right ]$ at lower temperatures \cite{ogasawara2021}.
The insulating state at low $T$'s has been interpreted as the property of a small semiconducting-like band gap \cite{yogeshsingh2009,saparov2013b}.
However, the Seebeck coefficient ($S$) of \bmb{} approaches zero as $T \to 0$ [Fig.~\ref{fig:dos}(e)].
The contrast behaviors of $\sigma(T)$ and $S(T)$ demonstrate an AL of finite DOS at $E_{\mathrm{F}}$ \cite{ [{See section 2.13 in }] mott2012}.

The insulating behavior of \bmb{} is effectively suppressed by $\vec{H}_{ab}$'s [Fig.~\ref{fig:dos}(d)].
The delocalization effect is isotropic with respect to the direction of $\vecHab$ in the crystallographic $ab$-plane \cite{ogasawara2021,huynh2019}.
Meanwhile, magnetic fields parallel to the AF axis ($\vec{H}_c$) result in a much smaller effect \cite{huynh2019,ogasawara2021,jansa2021}.
The in-plane magnetic field $\vecHab$ is thus the tuning parameter for the AL transition in \bmb{}.

\begin{figure}
  \includegraphics[width = .48\textwidth]{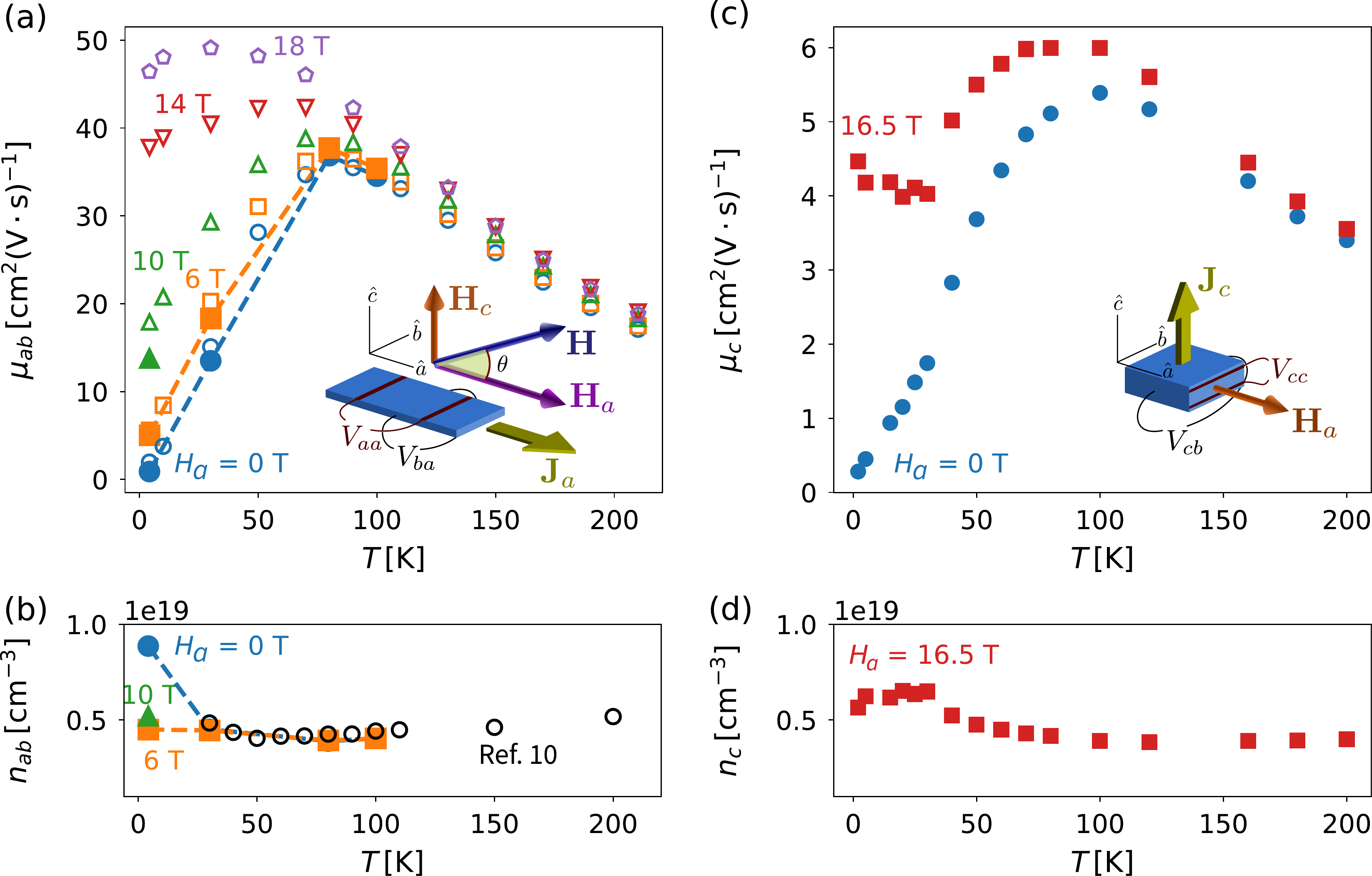}  
  \caption{(Color online)
    (a) Effects of $H_x$ on temperature dependencies of carrier mobility in the ab-plane ($\mu_{ab}$) estimated from the crossed-field Hall measurements \cite{Note1}.
    (b) Carrier number estimated from the high-$H_c$ Hall coefficient.
    The black empty circles represent the carrier estimated from the single field transverse Hall measurement published in Ref.~\onlinecite{ogasawara2021}.
    (c) Temperature dependencies of carrier mobility along the $\hat{c}$-axis ($\mu_c$) estimated for $H_a = \SI{0}{\tesla}$ and $\SI{16.5}{\tesla}$.
    (d) Carrier number estimated from the high-$H_a$ Hall coefficient.
    More details for the Hall measurements can be found in the Supplementary Information (SI) \cite{Note1}.
  }
  
  \label{fig:hall}
\end{figure}

\paragraph{Hall effects}---~%
\label{sec:hall}
In an AL state and/or in the vicinity of an AL transition, the Hall effect is a complex quantity that remains not well-understood \cite{friedman1971,friedman1977,arnold1974,fleishman1980a,mott1980,mott2012}.
In \bmb{}, the ability to accurately control the AL transition in the same sample via the $\vecHab$ provides a good opportunity to study the interesting topic.
We employed a crossed-field measurement method to investigate the current-in-plane ({\itshape cip}) Hall effect \cite{Note1}.
As schematically shown by the inset of Fig.~\ref{fig:hall}\,(a), $\vecHc$ is the probing field that helps measuring the transverse Hall response from the electronic state that is \emph{tuned by} $\vecHa$.
The effect of $\vecHa$ on the Hall resistivity $\rho_{xy}(\Hc)$ was investigated by rotating the sample in different static $\vecHa$'s \cite{Note1}.
The crossed-field measurements revealed tuning effects for $\vecHa$ on $\rho_{ba}$ \cite{Note1}.
At low $\Ha$'s, the $\rho_{ba}(\Hc)$ curves show a non-linearity that is related to VRH transport regime \cite{mott2012}.
In the metallic state at high $H_{a}$'s and/or high $T$'s, the non-linearity disappears and $\rho_{ba}$ returns to a simple linear line.

The crossed-field Hall measurements also allowed us to estimate the effects of $\vecHa$ on the carrier number and mobility.
Fig.~\ref{fig:hall}(a) shows that the in-plane mobility $\mu_{ab}$ displays a broad peak corresponding to the MIT at $\tmin$ and then decreases to zero when entering the VRH regime.
At high $H_{ab}$, $\mu_{ab}$ is metallic down to low $T$s.
The carrier number $n_{ab}$ is not affected by both $H_{a}$ and $T$ [Fig.~\ref{fig:hall}\,(b)].

A similar effects of $H_{ab}$ can be obtained from the out of plane Hall effect [inset of Fig.~\ref{fig:hall}(c)], in which $\vecHa$ takes a dual role of being the Hall and the tuning fields \cite{Note1}.
As shown in Fig.~\ref{fig:hall}(c), $\vec{H}_a$ effectively removes the insulating behavior at zero magnetic field of the interlayer mobility $\mu_c$.
On the other hand, the interlayer carrier number $n_c$, estimated at $H_x \geq \SI{15}{\tesla}$, does not change with $T$ [Fig.~\ref{fig:hall}(d)].
These deduced $\mu_{ab}$ and $\mu_c$ corroborate a clear delocalizing effect of $\vecHab$ on the $d_{xy}$ holes.

\paragraph{Scaling analyses}---~%
\label{par:scaling}%
Since the localization transition in the $d_{xy}$-derived band is tunable via $\vecHab$, one may anticipate that at $T = 0$ there exists a quantum critical point $H_{\mathrm{cr}}$ separating the metallic and insulating phases.
Using the scaling theory for the localization in both metallic and insulating regions \cite{wegner1976,belitz1994}, we investigated the criticality of the in-plane conductivity in the window $\SI{0.5}{\kelvin} \leq T \leq \SI{1}{\kelvin}$ by tuning $\Ha$.

\begin{figure}
  \centering  
  \includegraphics[width = \linewidth]{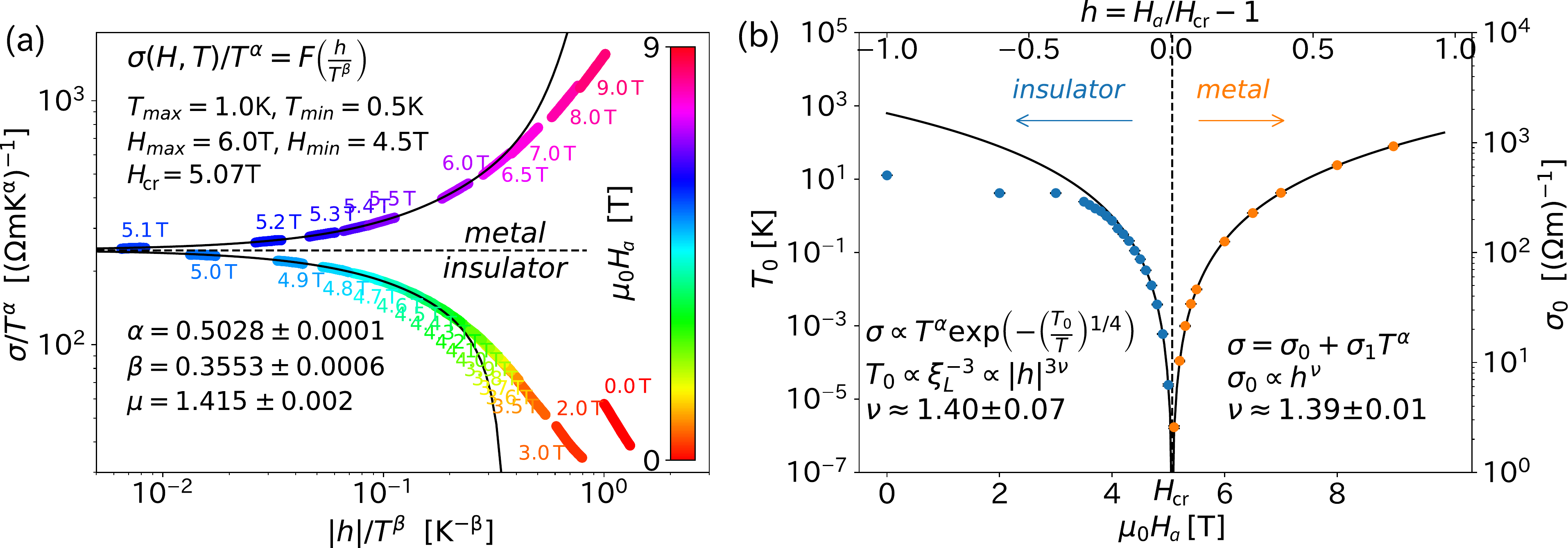}
  \caption{(Color online) Scaling properties of the MIT.
    (a) Finite-temperature scaling properties of $\sigma(T, H_a)$.
    The solid circles the represent the $\sigma(T)$ data measured at $H_a$'s as shown the texts and the color bar.
    The solid lines are the fit using a polynomial expansion of $\mathcal{F}$.
    (b) Scaling properties of $\sigma_0$ in metallic region and $T_0$ in the VRH insulating regime.
    The solid lines are power law fittings.
    $T_{max}$ ($H_{max}$) and $T_{min}$ ($H_{min}$) in (a) indicate the fitting range in $T$ ($H_a$).
      The scaling formulae for each scaling are also shown.
  }
  \label{fig:scaling}
\end{figure}

At finite temperatures, the in-plane conductivity measured under various $H_a$'s should obey the rule $\sigma(h,T) = T^\alpha\mathcal{F}\left(h/T^\beta\right)$. 
Here, $h = H_{ab}/H_{\mathrm{cr}}- 1$ is the dimensionless distance of magnetic field from the critical field $H_{\mathrm{cr}}$, and $h/T^\beta$ is the only variable of the scaling function $\mathcal{F}$.
At the vicinity of $H_{\mathrm{cr}}$, the localization (correlation) length $\xi$ diverges as $|H_{ab}- H_{\mathrm{cr}}|^{-\nu}$.
The critical exponents $\alpha$ and $\beta$ are related to the correlation length exponent $\nu_1$ and the dynamical exponent $z$ by $\alpha = (d-2)z^{-1} = z^{-1}$ and $\beta = (z\nu_1)^{-1}$, in which the $d = 3$ is the dimensionality of system.
Fig.~\ref{fig:scaling}(a) shows that the best scaling is achieved with $\alpha \approx 0.503$ and $\beta \approx 0.355$, corresponding to $z = \alpha^{-1} \approx 1.988$ and $\nu_1 = \alpha\beta^{-1} \approx 1.415$, by a polynomial fitting near $h = 0$ \cite{m.itoh2004}.
The fitting also yields the critical field $H_{ab}^{\mathrm{cr}} \approx \SI{5.07}{\tesla}$ \cite{Note1}.

On the other hand, at $T = 0$, the two sides across $h = 0$ exhibit entirely different states.
For the metallic state at $h > 0$, the zero temperature conductivity $\sigma_0$ is positive and finite, however $\sigma_0 < 0$ in the insulating region, $h < 0$.
We thus employed different approach for each region.
For $h>0$, the metallic conductivity generally obeys $\sigma_0 (h) \propto h^\mu$.
The critical exponent is related to $\nu$ by the definition $\mu = (d-2)\nu_2 = \nu_2$.
We estimated $\sigma_0$ from the power law fitting $\sigma(T, H_a) = \sigma_0(H_a) + aT^q$ at low temperatures; $q$ takes a value close to the inverse of the dynamical exponent in the finite temperature scaling, i.e. $q \approx 1/z \approx 0.5$ \cite{belitz1994,wegner1976}.
The fitting of obtained $\sigma_0$'s to the critical law $\sigma_0 \propto h^{\nu_2}$ (black line in the metallic region of Fig.~\ref{fig:scaling}(b)) yields $\nu_2 \approx 1.39$.

As shown in Fig.~\ref{fig:scaling}(b), in the insulating side ($h < 0$) at $T = \SI{0}{\kelvin}$, the scaling behavior appears in the $\Ha$-dependence of $T_0$ extracted from the modified VHR law $\sigma(T) \propto T^{1/2}\exp\left [ - ( T_0 / T)^{1/4} \right ]$ \cite{m.itoh2004}.
As a result of the small $T_0$ near $h=0$, the $T$ dependence of pre-factor is taken into account \cite{Note1}.
By definition, $T_0 \propto \xi^{-3}$, and $\xi$ diverges as $|h|^{-\nu_3}$ as $h \to 0$.
A $T_0 \propto |h|^{3\nu_3}$ fit (black line) yields $\nu_3  = 1.40$.
The deviations from the fittings at $H_{ab} < \SI{3}{\tesla}$ in Fig.~\ref{fig:scaling}(a) and (b) are mainly due to the effect of a an additional positive magnetoresistance \cite{huynh2019,ogasawara2021} that is not included in the scaling theory. 

The obtained values $\nu_i \approx 1.4$ points to an AL transition that belongs to the 3D unitary universality class, for which a $\nu^{\mathrm{cal}} \approx 1.4$ was obtained by simulations \cite{kawarabayashi1998,wang2021a}.
This is consistent with the absence of time-reversal symmetry caused by both the tuning parameter $h$ and the G-AF order.
The value $\nu\approx 1.4$ is also far from the critical exponent of an Anderson-Mott transition $\nu = 0.33$ for Cr-doped V$_2$O$_3$, and $\nu z = 0.56$ as obtained from theoretical calculations) \cite{limelette_universality_2003,terletska_quantum_2011}.
Randomness plays a decisive role and electron-electron interactions are of less importance despite the strong electron correlations of the G-AF background.

\begin{figure}
  \includegraphics[width =\linewidth]{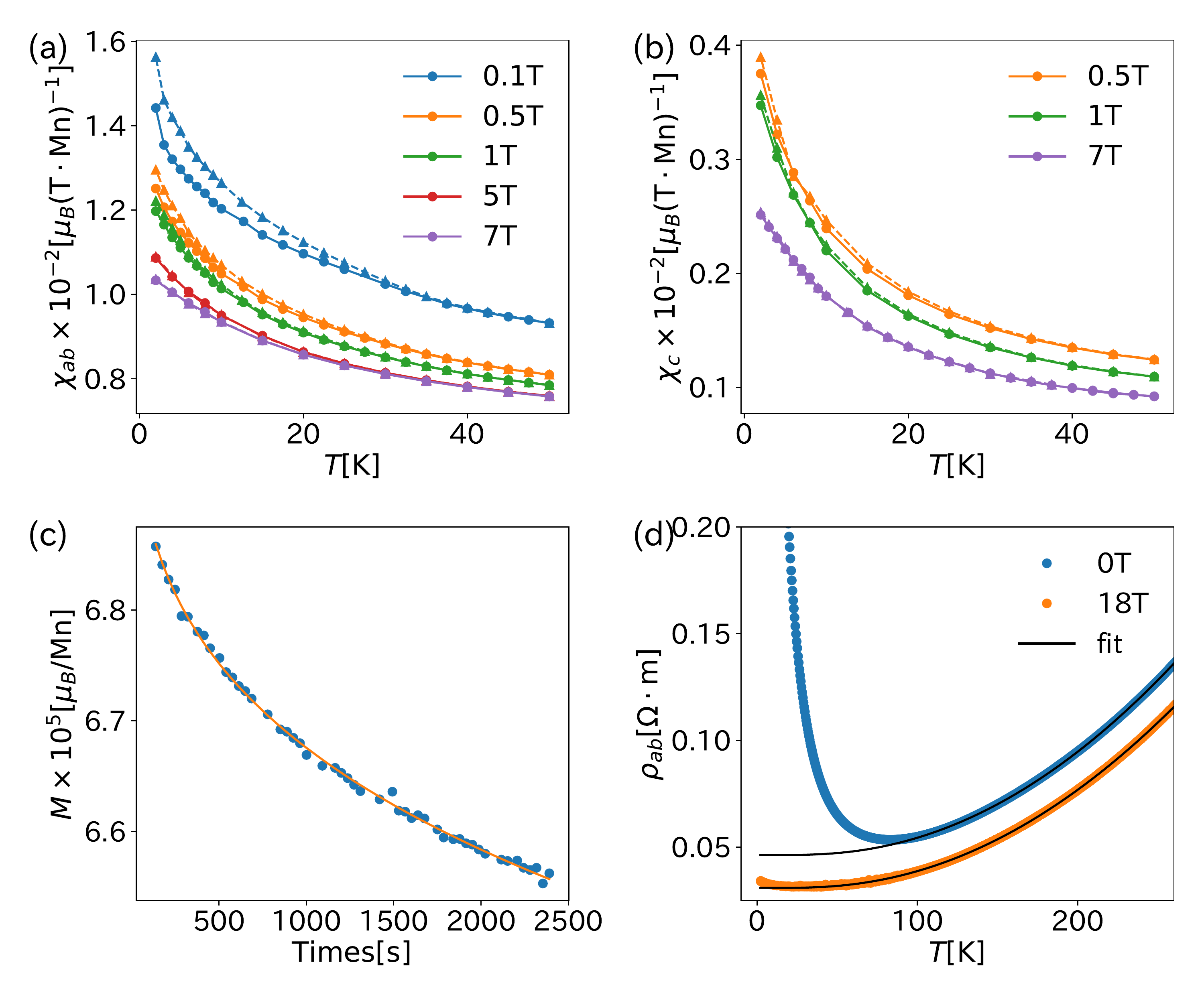}
  \caption{(Color online) (a) and (b) Temperature dependence of magnetic susceptibility of field cool and zero field cool measured at different values of $H_{ab}$ and $H_{c}$, respectively.
    The circle and triangle indicate zero field cool and field cool, respectively.
    (c) Time dependence of magnetization after turning off $H_{a} = \SI{0.1}{\tesla}$.
    (d) Temperature dependencies of resistivity measured under $H_{a} = \SI{0}{\tesla}$ and $\SI{18}{\tesla}$.
    The black lines show the fits using a phenomenological rule for half-metals \cite{bombor2013}.
  }
  \label{fig:magnetization}
\end{figure}

\paragraph{A hidden spin glass-like transition}---~
\label{sec:hidden-spin-glass}
By carefully reexamining the magnetization \cite{Note1}, we found characteristics of a spin glass (SG) -like state  that had been hidden by the G-AF.
Fig.~\ref{fig:magnetization}(a) shows that the zero field cool (ZFC) and field cool (FC) curves of the in-plane magnetic susceptibility $\chi_{ab}$ split at the irreversible temperature $T_f \approx \SI{30}{\kelvin}$.
The splitting $(\chi^{\mathrm{FC}}-\chi^{\mathrm{ZFC}})$ is about $1.2\times 10^{-3}\mu_B/\mathrm{Mn/T}$ at $\Hab = \SI{0.1}{\tesla}$.
The splitting of the out-of-plane magnetic susceptibility $\chi_{c}$ shown in Fig.~\ref{fig:magnetization}(b) is approximately three times smaller than the case of $\chi_{ab}$.
This magnitude is much smaller than that of typical SG systems \cite{binder1986}, but it is of the same order with that of the SG coexisting with AF in $\mathrm{Fe_{1/3}NbS_2}$ \cite{maniv2021a}.
The time decay of the remnant magnetization can be fitted to a stretched exponential law $M(t) = M_0\exp{(- a t^{1-n})}$ [Fig.~\ref{fig:magnetization}(c)].
The fit yields $1-n \approx 0.257$, which is close to the theoretical value for a canonical SG, $1 - n  = 1/3$ \cite{chu_dynamic_1994}.
The irreversibility decreases quickly with increasing $\Hab$, and completely vanishes at $\Hab \gtrapprox H_{\mathrm{cr}} (\approx \SI{5}{\tesla})$.
This observation may indicate a connection between the $\vecHab$-tuned electronic delocalization of the $d_{xy}$ holes and the SG.

A question arises on the magnetic properties in the metallic regime.
Since a direct measurement for the $d_{xy}$ magnetism at high $H_{ab}$ is difficult due to the smallness of SG-like magnetization,
we tried an approached using the transport properties.
As shown in Fig.~\ref{fig:magnetization}(d), the temperature dependence of resistivity measured at $H_{ab} = \SI{18}{\tesla}$ agrees very well with the expectations for the FM half-metals, i.e. $\rho(T) = \rho_0 + aT^2\exp(-\Delta/T)$, given $\Delta = \SI{8}{\milli \electronvolt}$ as the spin gap.
This high-$H_{ab}$ metallic behavior strongly resembles that of the heavily hole-doped BaMn$_2$As$_2$ \cite{ pandey2013b}. 
We therefore surmise that that the AL accompanied by SG-like state eventually evolve to an intriguing half-metallic FM at the $h \gg 0$ limit \footnote{A more rigorous derivation for the origin of SG and its relationship to the AL transition is difficult to achieve due to the smallness of the SG magnetization.
  We cannot totally rule out the possibility that SG originates from a small number of unknown impurities.
}.

\paragraph{Discussion and Summary}---~%
\label{sec:discussion-summary}
The critical exponent $\nu \approx 1.4$ helps us to place the $\vecHab$-induced AL transition in the unitary universality class.
An unitary, i.e., broken time-reversal symmetry, AL can originate from the randomness in the phase of the electronic hopping amplitude that is introduced by an internal random gauge field, i.e., the magnetic field \cite{kawarabayashi1998,
  [{In the other case, the AL stems from the disorders of the crystal potential, and an external magnetic field can merely weaken it via breaking the time-reversal symmetry, causing a reduction of about $0.1$ in the value $\nu$. See }]slevin1997}.
Thus, a sufficiently strong external magnetic field can induce a delocalization by suppressing the randomness of the internal vector potential.
Because the AL transition is tunable solely via the perpendicular $\vecHab$ \cite{huynh2019,ogasawara2021}, the random gauge field causing it has a close connection to the fluctuations of the Mott-like AF.
It has been known that a perpendicular field $\vecHab$ can suppress the G-AF fluctuations effectively whereas a parallel $\vec{H}_c$ cannot \cite{li2012}.

In Fig.~\ref{fig:dos}(b), the large energy separation between the $d_{xy}$-derived band and the $3d$ G-AF states suggests that the hoping of the quasi-free holelike charges at $\EF$ can cause only infinitesimal perturbations to the latter.
On the other hand, from the viewpoint of spin degrees of freedom, the quasi-static AF spins and the spins of the $d_{xy}$ holes are still coupled via spin-spin interactions \cite{takaishi_localization_2018}.
AF spin fluctuations hence are inherited as phase randomness of the hopping amplitudes, and an AL of charge can be expected as proposed theoretically \cite{takaishi_localization_2018}.
The frozen spins of the $d_{xy}$ holes in the AL state can be the origin of the observed SG-like magnetization \cite{*[{The spins at the $\EF$ should have their own AF or FM interactions, which may competes with the interaction with the G-AF. See }] [{, }] ,blundell2001}.

The quasi-one particle characteristics of the $\vecHab$-tuned AL amid strong correlation can be viewed as a consequence of the intriguing multiorbital electronic structure.
The itinerancy of the $d_{xy}$-derived band has been shown to be robust even against large $U$ \cite{craco2018}, and this supports a picture having different strengths of charge and spin couplings between the itinerant $d_{xy}$ and the deep AF states.
The current picture of the AL that stems from the interactions between $d_{xy}$ and AF spins also bears some similarities to a recent theoretical model, in which a quasi-many body localization is shown to arise from the coupling between coexisting fast and slow particles \cite{yao2016}.

\section*{Acknowledgements}
Measurements of transport properties and magnetism were carried out at Advance Institute for Materials Research and High Field Laboratory for Superconducting Materials at Tohoku University, and at Center for Advanced High Magnetic Field Science at Osaka University via Visiting Researcher's Program of the Institute for Solid State Physics, the University of Tokyo.
  We thank K. Nomura, K. Ogushi, T. Sato, T. Aoyama, K. Igarashi, H. Watanabe, Y. Yanase, T. Urata, and T. Arima for fruitful discussions.
  This work was supported by the CREST project ``Thermal Management'', and by the Grant-in-Aid for Scientific Research on Innovative Areas ``J-Physics'' (Grant No.18H04304), and by JSPS KAKENHI (Grants No. 18K13489, No. 18H03883, No. 17H045326, and No. 18H03858).
  T.O. thanks the financial supports from the International Joint Graduate Program in Materials Science (GP-MS) of Tohoku University.
  This research was partly made under the financial support by the bilateral country research program of JSPS between AIMR, Tohoku University and Jozef Stefane Institute, Slovenia.
  This work was also supported by World Premier International Research Center Initiative (WPI), MEXT, Japan.
  DA acknowledges the financial support of the Slovenian Research Agency through BI-JP/17-19-004, J1-9145, and J1-3007 grants.


\bibliographystyle{apsrev4-1}
\bibliography{BaMn2Pn2_related.bib}

\end{document}


\title{Magnetic-field induced Anderson localization \\in orbital selective antiferromagnet \bmb{}\\-- Supplementary Information}

\author{Takuma Ogasawara}
\affiliation{Department of Physics, Graduate School of Science, Tohoku University, \\
  6-3 Aramaki, Aoba, Miyagi, Japan}

\author{Kim-Khuong Huynh}
\email{huynh.kim.khuong.b4@tohoku.ac.jp}
\affiliation{Advanced Institute for Materials Research (WPI-AIMR), Tohoku University, 1-1-2 Katahira, Aoba, Sendai, Miyagi, Japan}

\author{Stephane Yu Matsushita}
\affiliation{Advanced Institute for Materials Research (WPI-AIMR), Tohoku University, 1-1-2 Katahira, Aoba, Sendai, Miyagi, Japan}

\author{Motoi Kimata}
\affiliation{Institute for Material Research, 1-1-2 Katahira, Aoba, Sendai, Miyagi, Japan}

\author{Time Tahara}
\affiliation{Center for Advanced High Magnetic Field Science, Graduate School Science, Osaka University,
  1–1 Machikaneyama, Toyonaka, Osaka, Japan}

\author{Takanori Kida}
\affiliation{Center for Advanced High Magnetic Field Science, Graduate School Science, Osaka University,
  1–1 Machikaneyama, Toyonaka, Osaka, Japan}

\author{Masayuki Hagiwara}
\affiliation{Center for Advanced High Magnetic Field Science, Graduate School Science, Osaka University,
  1–1 Machikaneyama, Toyonaka, Osaka, Japan}

\author{Denis Ar\v{c}on}
\affiliation{Faculty of Mathematics and Physics, University of Ljubljana,
  Jadranska c. 19, 1000 Ljubljana, Slovenia}
\affiliation{Jozef Stefan Institute,
  Jamova c. 39, 1000 Ljubljana, Slovenia}

\author{Katsumi Tanigaki}
\email{katsumi.tanigaki.c3@tohoku.ac.jp}
\affiliation{Advanced Institute for Materials Research (WPI-AIMR), Tohoku University, 1-1-2 Katahira, Aoba, Sendai, Miyagi, Japan}
\affiliation{BAQIS, Bld. 3, No.10 Xibeiwang East Rd., Haidian District, Beijing 100193, China}

\maketitle

\label{sec:authorlist}

\section{Experimental details}
\label{sec:experiments}
We synthesized \bmb{} single crystals via a self-flux method in which Bi was used as the flux.
Magnetotransport properties of the crystals were measured using a Quantum Design Physical Properties Measurement System (PPMS) for $H \leq \SI{9}{\tesla}$ at Advanced Institute for Material Research, Tohoku University and a cryostat equipped with a $\SI{18}{\tesla}$ superconducting magnet at High Field Laboratory for Superconducting Magnet, Institute for Material Research, Tohoku University.
Measurements of magnetoresistance under magnetic fields up to $\SI{50}{\tesla}$ were carried out at Center for Advanced High Magnetic Field Science, Graduate School of Science, Osaka University. We used standard four-point-probe to measure the resistivity.
Angular dependencies of magnetoresistance and Hall effect were measured with the help of a dual axis rotator developed by Dr. Motoi Kimata.
We collected the angular dependence of Hall effect by rotating the samples under various static magnetic fields while continuously measuring the Hall voltage (see Sec.~\ref{sec:HallSM}).
Magnetic properties of the single crystals were measured using a Quantum Design Magnetic Properties Measurement Systems (MPMS).
In order to measure the Seebeck coefficient, a two-heater-two-thermometer method was used, the details of which will be published elsewhere.

\section{First principle calculation based on Density Functional Theory}
\label{sec:dft}

We used Density Functional Theory (DFT) codes WIEN2k and Quantum Espresso to calculate the electronic structure of \bmb{} \cite{blaha_wien2k_2020,giannozzi2009a}.
We used the lattice parameters either taken from the experiment \cite{saparov2013b} or by structural optimization with the assumption of G-type antiferromagnetic order (AF).
Both calculations show very similar results.
In addition, the band structures obtained from both codes show also very little differences.
Generalized Gradient Approximation (GGA) was used to calculate the exchange correlation energy.
The number of $k$-points was usually chosen to be $1000$ for self-consistent calculation and $8000$ for non-self-consistent calculation in the calculations of the density of states.

\begin{figure}
    \begin{minipage}[c]{0.70\linewidth}
    \centering
    \includegraphics[width=0.95\linewidth]{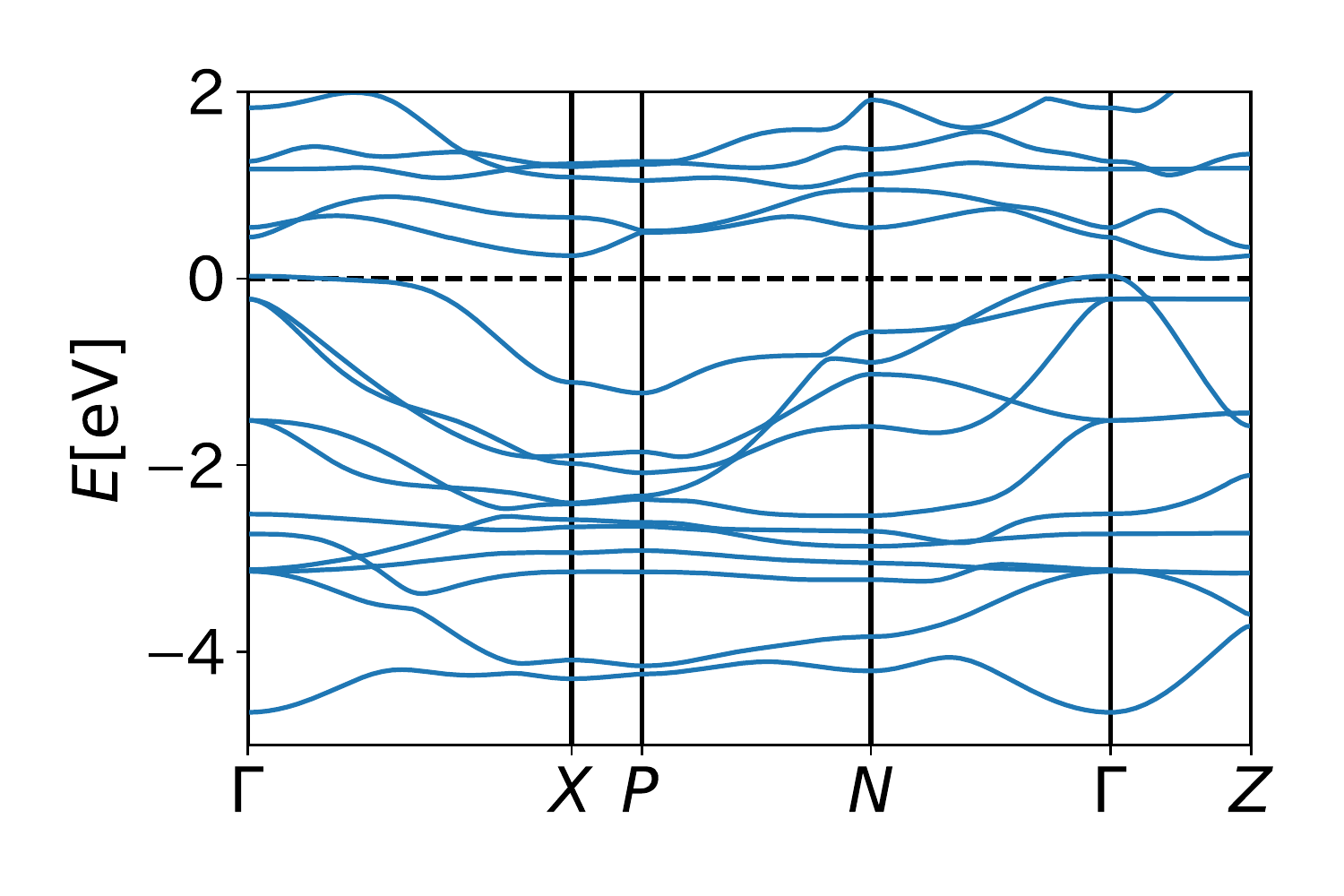}
    \end{minipage}
    \begin{minipage}[c]{0.25\linewidth}
    \centering
    \includegraphics[width=0.95\linewidth]{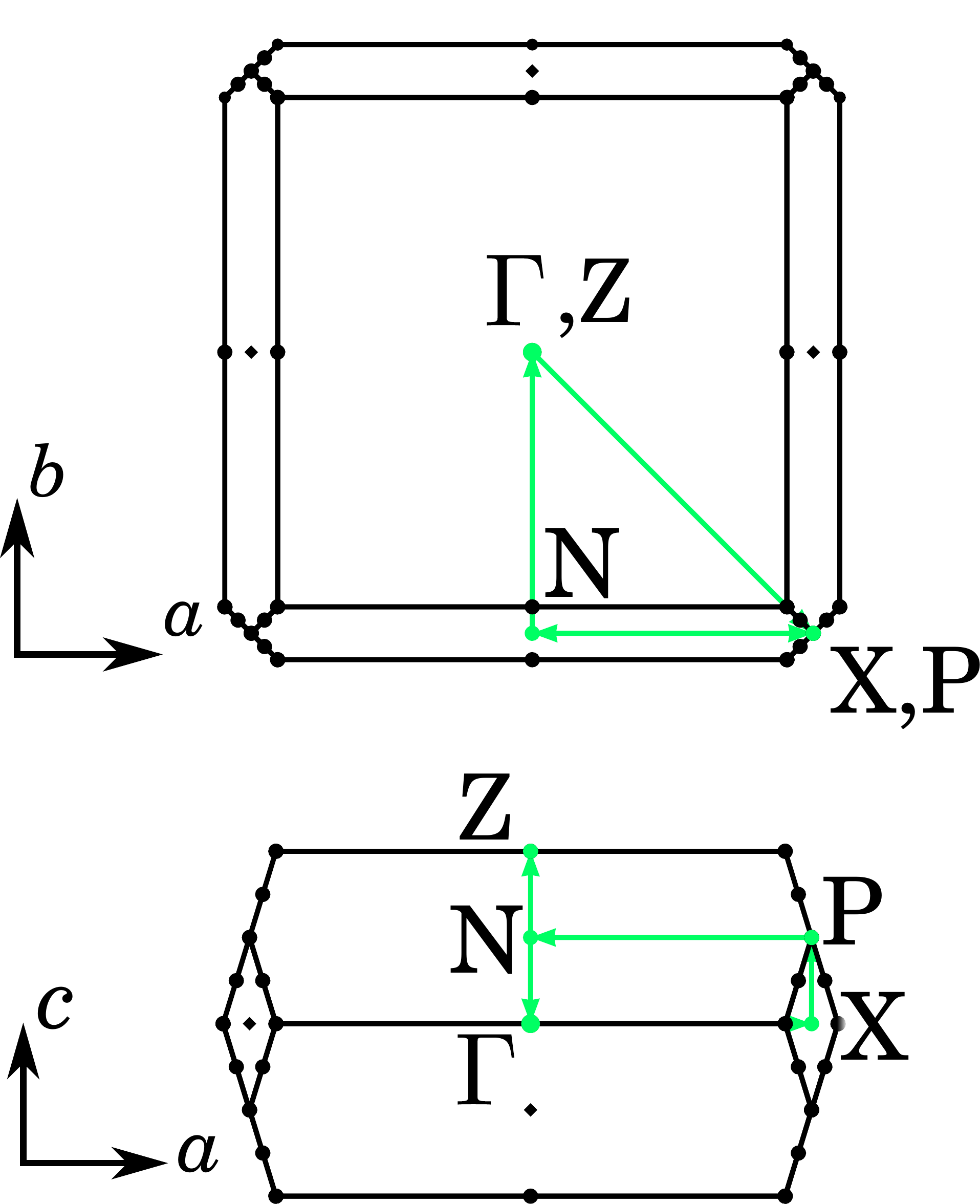}
    \end{minipage}
    \caption{{\itshape Left:} The spin-resolved band structure of antiferromagnetic \bmb{}.
      Only spin-up band is shown.
      {\itshape Right:} Highly symmetrical lines in the first Brillouin zone.
    }      
    \label{fig:band_normal}
\end{figure}

Fig.~\ref{fig:band_normal} shows the band dispersion along the high symmetry directions in the first Brillouin zone.
The valence bands in the energy range from $\SI{-5}{\electronvolt}$ to $\SI{0}{\electronvolt}$ (Fermi level, $\EF$) consist of eleven bands for five $3d$-electrons from Mn (in each AF sublattice) and six $p$-electrons from two Bi atoms in the unit cell.
The entanglements of the bands may favor the hybridizations between different orbitals.

\begin{figure}[hpbt]
    \centering
    \includegraphics[width=.7\linewidth]{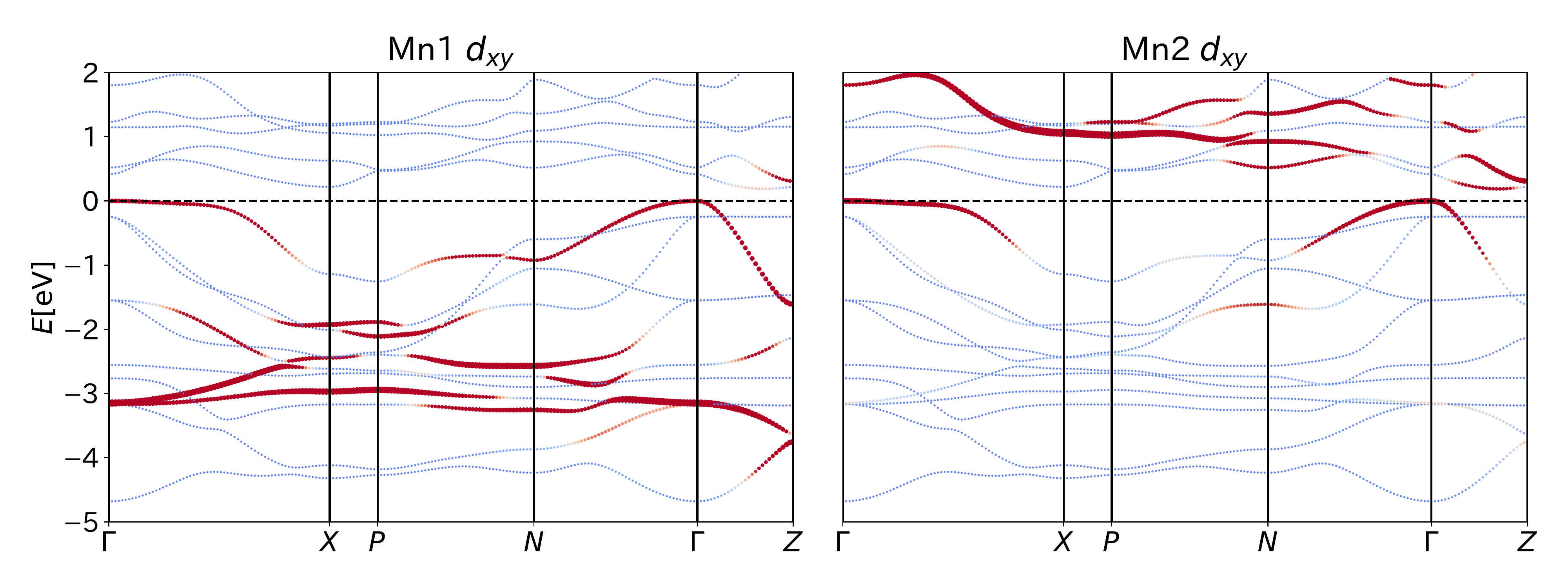}
    \includegraphics[width=.7\linewidth]{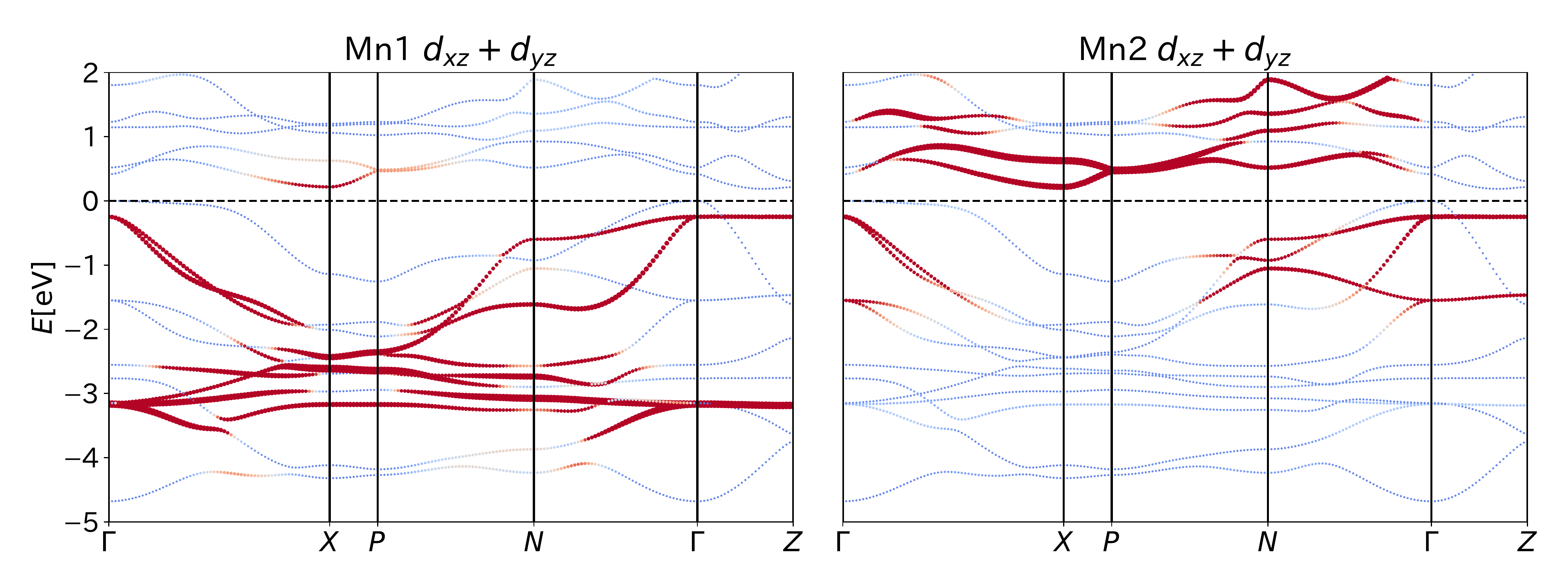}
    \includegraphics[width=.7\linewidth]{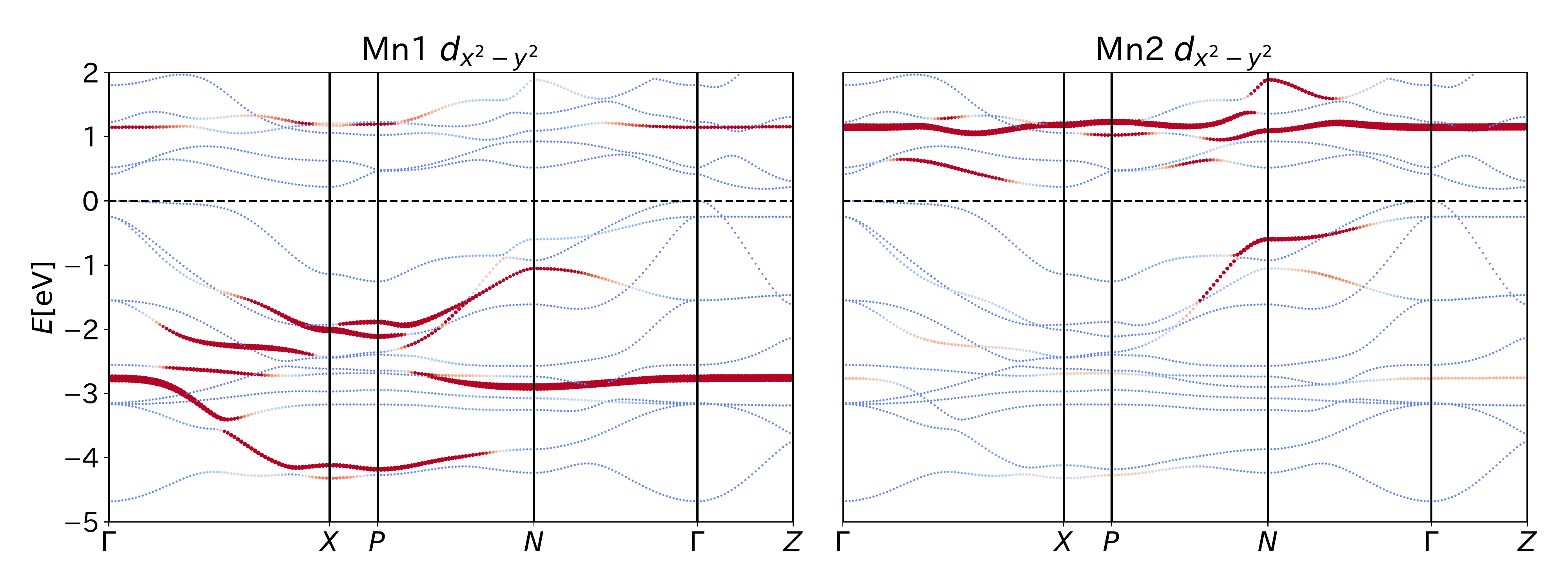}
    \includegraphics[width=.7\linewidth]{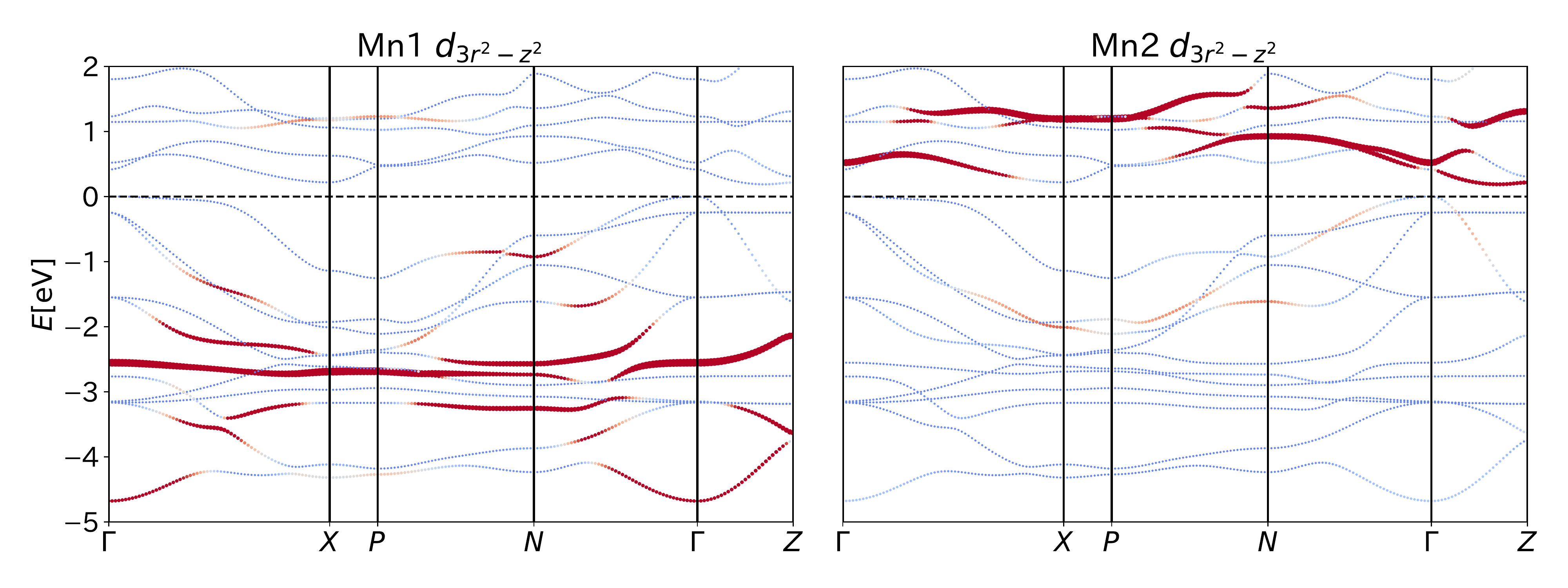}
    \includegraphics[width=.7\linewidth]{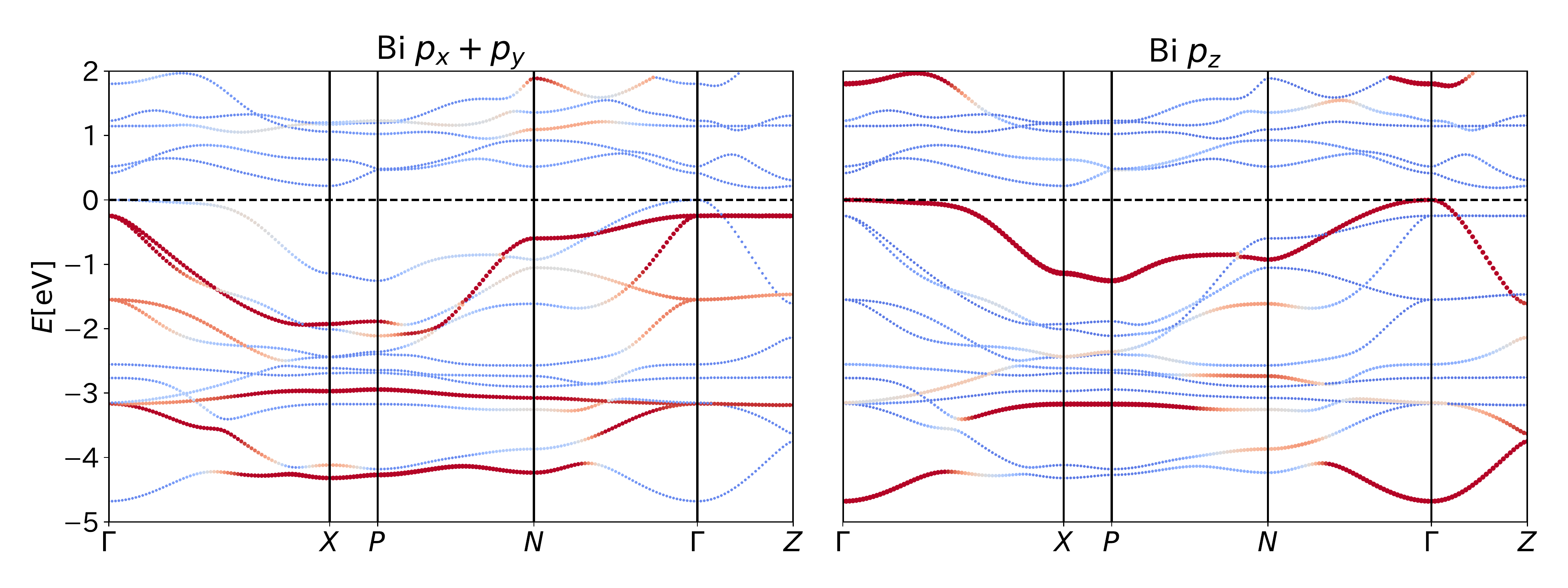}
    \caption{The band structure of GGA calculation for \bmb{} with the orbital projection against $d$-orbitals at a Mn site and $p$-orbitals at a Bi site, indicated by the size of markers. Mn1 and Mn2 refer to Mn sites with spin up (majority state) and spin down (minority state) sublattices, respectively.}
    \label{fig:band_heavior}
\end{figure}

Fig.~\ref{fig:band_heavior} shows the spin-polarized heavier plots for the weights of $d$-orbitals at Mn sites and $p$-orbital at Bi sites.
A clear orbital-selective AF can be seen.
Whereas a clear spin polarization of $d$-orbitals can be seen at $E < \SI{-2}{\electronvolt}$, the contributions of both spin-up and spin-down components are almost equivalent in the energy window  $\SI{-2}{\electronvolt} \geq E \geq \SI{0}{\electronvolt}$.
The spin polarizations of individual each orbitals also vary.
The $d_{x^2 - y^2}$ and $d_{3r^2-z^2}$ orbitals contributes only to the deep bands and are fully spin-polarized.
On the other hand, the weights of the $d_{xz} + d_{yz}$ and $d_{xy}$ orbitals are large in both deep, well-polarized and shallow, non-polarized bands.
\emph{The electronic structure near the Fermi level then consists of a mixture of both up and down spin components and is almost entirely made of $d_{xy}$ orbital.}
The partial spin polarization of the Mn state deduced from these band calculations is consistent with the value of measured AF magnetic moment at Mn site, which is approximately $3.89\,\mu_B$ per Mn site \cite{calder_magnetic_2014}.

\begin{figure}[hpbt]
    \centering
    \includegraphics[width=0.5\linewidth]{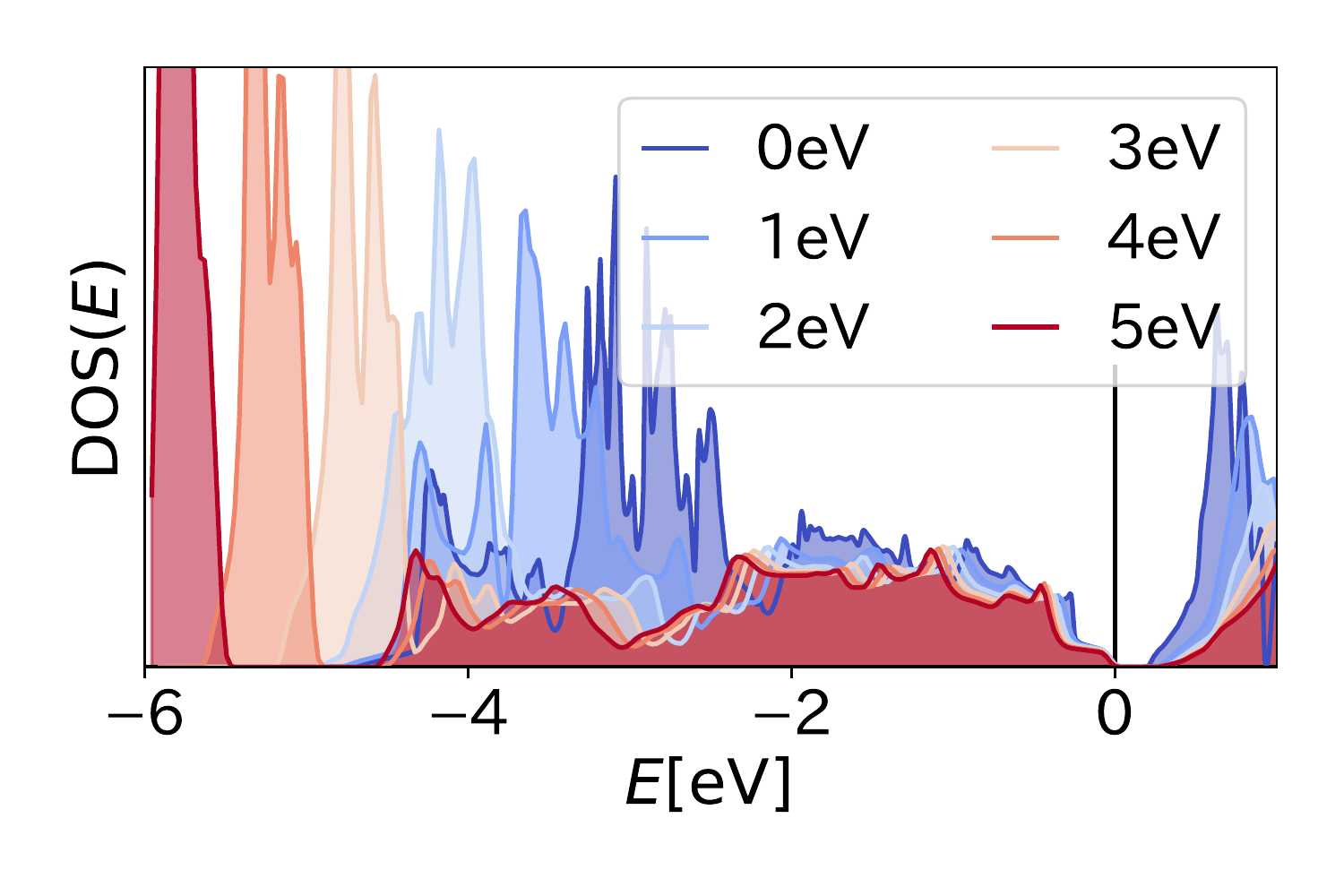}
    \caption{Density of states of \bmb{} calculated by DFT+U with various strengths of Hubbard parameter $U$ on Mn.}
    \label{fig:dos_us}
\end{figure}

We employed DFT+U calculations to investigate the robustness of orbital selective state with respect to electron correlation $U$.
As shown in Fig.\ref{fig:dos_us}, the AF-polarized states are pushed down by increasing $U$.
On the other hand, the non-spin polarized bands at $E > \SI{-2}{\electronvolt}$ are almost unaffected by $U$.
The DFT and DFT+U results shown here are similar to those obtained from a Dynamic Mean Field Theory calculations for \bma{} \cite{craco2018}.

\section{Crossed-field Hall effects}
\label{sec:HallSM}
\subsection{General considerations}
\label{sec:hall-settings-general}
At zero in-plane magnetic field ($\vecHab = 0$), \bmb{} exhibit an Anderson localization (AL) at low temperatures.
A strong $\vecHab$ that is \emph{perpendicular} to the antiferromagnetic sublattice can tune the insulating state back to metallic, as depicted in Fig.~1(d).
The scaling analyses shown in Fig.~3 of main text and in Sec.~\ref{sec:scalingSM} of the SI demonstrate that the $\vecHab$-tuned metal-insulator transition (MIT) can be understood as an unitary Anderson localization (AL) transition.

In an AL state and/or in the vicinity of an AL transition, the Hall effect is a complex quantity that remains not well-understood \cite{friedman1971,friedman1977,arnold1974,fleishman1980a,mott1980,mott2012}.
Since the AL transition in \bmb{} is controllable via $\vecHab$, a better tuning of the criticality within the same sample can be achieved.
Therefore, \bmb{} can be a good model material to study this interesting but rather complex topic.

\begin{figure}
  \includegraphics[width = \textwidth]{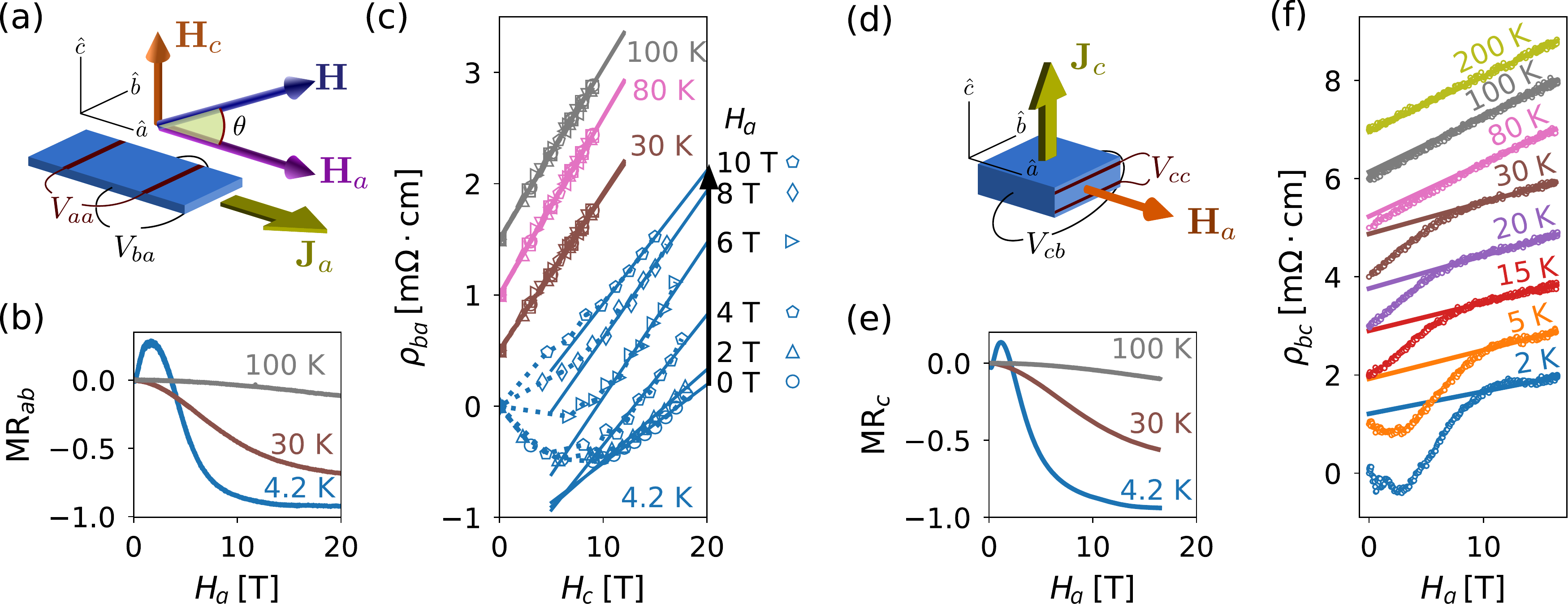}
  
  \caption{(Color online) --
    (a) Settings for the quasi crossed-field measurements of the current-in-plane ({\itshape cip}) transport properties.
    The $\hat{a}$-, $\hat{b}$-, and $\hat{c}$-axes are main crystallographic directions.
    (b) {\itshape cip}\,-MR as a function of the in-plane magnetic field $H_{a}$ at different temperatures (taken from \cite{ogasawara2021}).
    (c) The dependence of $\rho_{ba}$ on $H_c$ measured at different $H_a$'s and temperatures.
    The dotted lines are for guiding the eyes.
    The solid lines are linear fits.
    The curves measured at different temperatures are shown by different colors and shifted for clarity.
    At each temperature, each symbol shape represents the $\rho_{ba}(H_c)$ curve collected at a constant value of $H_a$, as annotated for $T=\SI{4.2}{\kelvin}$.
    The same symbol shape convention is used for the data at other temperatures.
    In the VRH-regime, $T=\SI{4.2}{\kelvin}$ and small $H_a$, $\rho_{ba}(H_c)$ is nonlinear.
    Strong $H_a$'s flatten the $\rho_{ba}(H_c)$ curves to a linear line.
    At high $T$'s, $\rho_{ba}(H_c)$ is linear and its slope does not change with $H_a$.
    (d) Settings of the current-out-of-plane ({\itshape coop}) measurements. 
    (e) {\itshape coop}\,-MR as a function of the in-plane magnetic field $H_{a}$ \cite{ogasawara2021}.
    (f) {\itshape coop}\,-Hall effects at different temperatures.
    The curves were shifted for clarity.
    The solid lines are linear fits.
  }
 
  \label{fig:hallSM}
\end{figure}

A few useful points can be gathered from the previous works on the Hall effects in an AL state \cite{mott1980,mott2012}.
At first, the sign of the Hall effect in the variable-range hopping regime (VRH) can be opposite to the carrier type, i.e., VRH holelike carriers can exhibit a positive Hall effect.
The VRH-induced anomalous sign of the Hall effect is consistent with our observations for \bmb{}.
As shown by Fig.~1(e) in the main text, the positive Seebeck coefficient indicates that carriers are indeed holelike.
On the other hand, Fig.~\ref{fig:hallSM}(c) shows that in the VRH regime, the transverse Hall effect ($H_a = \SI{0}{\tesla}$) exhibit a negative sign at low $H_c$ (see also Ref.~\onlinecite{ogasawara2021}).
Secondly, the Hall mobility of an AL system generally decreases with temperature \cite{arnold1974,mott1980,mott2012}.

Since a  strong $\vecHab$ delocalizes the holelike carriers, it can also remove the VRH-induced negative sign of the Hall effect.
Furthermore, in the metallic state at high $H_{ab}$, the Hall mobility should also increase with decreasing temperatures.
The enhancement of the mobility due to the delocalization effect is in sharp contrast to a Liftshitz transition induced by a Zeeman splitting; the latter is associated with a sharp change in carrier number.
It is thus instructive to study the influence of $\vecHab$ on the Hall resistivity $\rho_{ba}$.

Because the MIT is tunnable via $\vecHab$, two simultaneous magnetic fields are important to study the $\rho_{ba}$ of \bmb{}.
The first magnetic field is the \emph{tuning field} $\vecHab$, which is always perpendicular to the $\hat{c}$-axis.
$\vecHab$ plays the role of tuning \bmb{} from insulating to metallic states.
In order to minimize the Lorentz force exerted by $\vecHab$ on the electric current, one can align this tuning magnetic field along the electric current, $\vec{H}_a \parallel \vec{J}$.
In addition to $\vec{H}_a$, for the measurement of the Hall resistivity $\rho_{ba}$, one needs to apply a \emph{Hall field} that is perpendicular to the $ab$-plane, i.e. $\vec{H}_c$.
In the ideal case, $\vec{H}_a$ does not have any other effects than tuning of the electronic state of the sample, \emph{and $\vec{H}_c$ probes the Hall response from the electronic state that is defined by $\vec{H}_a$.}
Crossed-field Hall measurements was previously employed to study field-tuned quantum critical point in the heavy fermion system YbRh$_2$Si$_2$ \cite{Paschen2004}.

\subsection{Crossed-field Hall experiments for \bmb{}}
\label{par:Hall-settings}
A vector magnet can generate simultaneously two orthogonal magnetic fields; therefore it provides a direct approach to measure the crossed-field Hall effect.
The sample can be aligned so that the electric current is parallel to the longitudinal magnetic field, and its plane perpendicular to transverse field.
In such a setting, the longitudinal and transverse magnetic fields play the roles of the tuning and Hall fields ($\vec{H}_a$ and $\vec{H}_c$), respectively.
In the VRH-regime, the in-plane Hall resistivity ($\rho_{ba}$) of \bmb{} is nonlinear at $H_c \leq \SI{10}{\tesla}$.
On the other hand, the critical tuning field is $H_{{\mathrm{cr}}} \approx \SI{5}{\tesla}$, as shown in Fig.~3 of the main text.
In order to put \bmb{} completely into the metallic state, $H_a$ has to be far higher than $\SI{5}{\tesla}$.
In other words, the upper fields should higher than $\SI{10}{\tesla}$ for both longitudinal and transverse fields.

Other than using a vector magnet, we can approach the crossed-field measurement using a dual-axis rotator and a conventional superconducting magnet, as schematically shown in Fig.~\ref{fig:hallSM}(a).
The electric current $\vec{J}_a$ was applied along the crystallographic $\hat{a}$-axis, and the two Hall electrodes probe the transverse voltage drop $V_{ba}$ along $\hat{b})$-direction on the sample.
We first aligned the sample into the longitudinal condition, i.e. $\vec{J}_a$ is parallel to the external magnetic field $\vec{H}$.
The sample was then rotated about its $\hat{b}$-axis in the static magnetic field, and the angular dependencies of $\rho_{aa}$ and $\rho_{ba}$ were measured.
Although $\vec{H}$ can only take two opposite directions parallel to the $\hat{z}_{\mathrm{mag}}$-axis of the magnet, rotating the sample about its $\hat{b}$-axis is equivalent to rotating $\vecH$ in the $ac$ plane of the sample.
The dual-axis rotator allowed us to carry out fine alignments that helped keeping $\vecH$ within the $ac$-plane during the experiments.

Given $\theta$ as the angle between $\vecH$ and $\vec{J}_a$ and $H = |\vecH|$, the magnitudes of the tuning field $H_a$ and the Hall-field $H_c$ are,
\begin{subequations}
  \label{eq:rotate-field}
\begin{align}  
  H_a &= H\cos\theta\,,\\
  H_c &= H\sin\theta\,.
\end{align}  
\end{subequations}
The measurements of the angle-resolved $\rho_{ba}$ at multiple $H$'s thus provide information about the $H_a$-dependence of the Hall effect (see Eq.~\eqref{eq:rhoHxHz} below) and can be viewed as an equivalence of the crossed-field Hall method.

The symmetry constraint that a true Hall resistivity $\rho_{ba}$ is odd under reversing the direction of $\vecH$ is helpful for data analyses.
Therefore, at first, we measured the angular dependencies of $\rho_{ba}^{\mathrm{meas}}$ at a set of static field strengths $(H_1, H_2, \ldots, H_n)$.
Next, the angular dependencies were measured at $(-H_1, -H_2, \ldots, -H_n)$.
The true $\rho_{ba}$ was then calculated as,
\begin{align}
  \label{eq:rhoyx}
  \rho_{ba} (H_i, \theta) = \frac{\rho_{ba}^{\mathrm{meas}}(H_i, \theta) - \rho_{ba}^{\mathrm{meas}}(-H_i, \theta)}{2}\,.
\end{align}
Eq.~\eqref{eq:rhoyx} effectively eliminates possible contaminations from imperfect positions of Hall electrodes.
\emph{The obtained $\rho_{ba}(H, \theta)$ is thus the true Hall resistivity}.

In terms of Eqs.\,\eqref{eq:rotate-field},
\begin{align}
  \label{eq:rhoHxHz}
  \rho_{ba} (H_i, \theta) \equiv \rho_{ba} (H_{i,a}, H_{i,c})\,.
\end{align}
The experimentally obtained data array $(\rho_{ba}, H_a, H_c)$ contains the values of $\rho_{ba}$ at different $H_a$'s and $H_c$'s.
For instance, in order to plot the $H_c$-dependence of $\rho_{ba}$ at a constant $H_{k,a}$, one can select from the array all the $(\rho_{ba}, H_c)$ pairs at $H_{k,a}$ [Fig.~\ref{fig:rhoyx-fixedH}].
We performed the procedure described above at various temperatures, with $H$ at $ \SI{3}{\tesla},\SI{6}{\tesla}$, and from $\SI{6}{\tesla}$ to $\SI{18}{\tesla}$ at a $\SI{1}{\tesla}$ step.
The results are shown in Fig.~\ref{fig:hallSM}(c).

As we will discuss below, the quasi-crossed field Hall measurement has a few limitations that fortunately do not affect on our interpretations and conclusions.
On the other hand, this method provides a feasible approach to study the Hall effect near the $\vecHab$-tuned AL transition in a wide range of crossed-fields, which can be otherwise difficult if we employ a vector magnet.

\subsection{Upper limits for $(H_a,H_c)$}
\label{sec:}
\begin{figure}
  \centering
  \includegraphics[width = .45\textwidth]{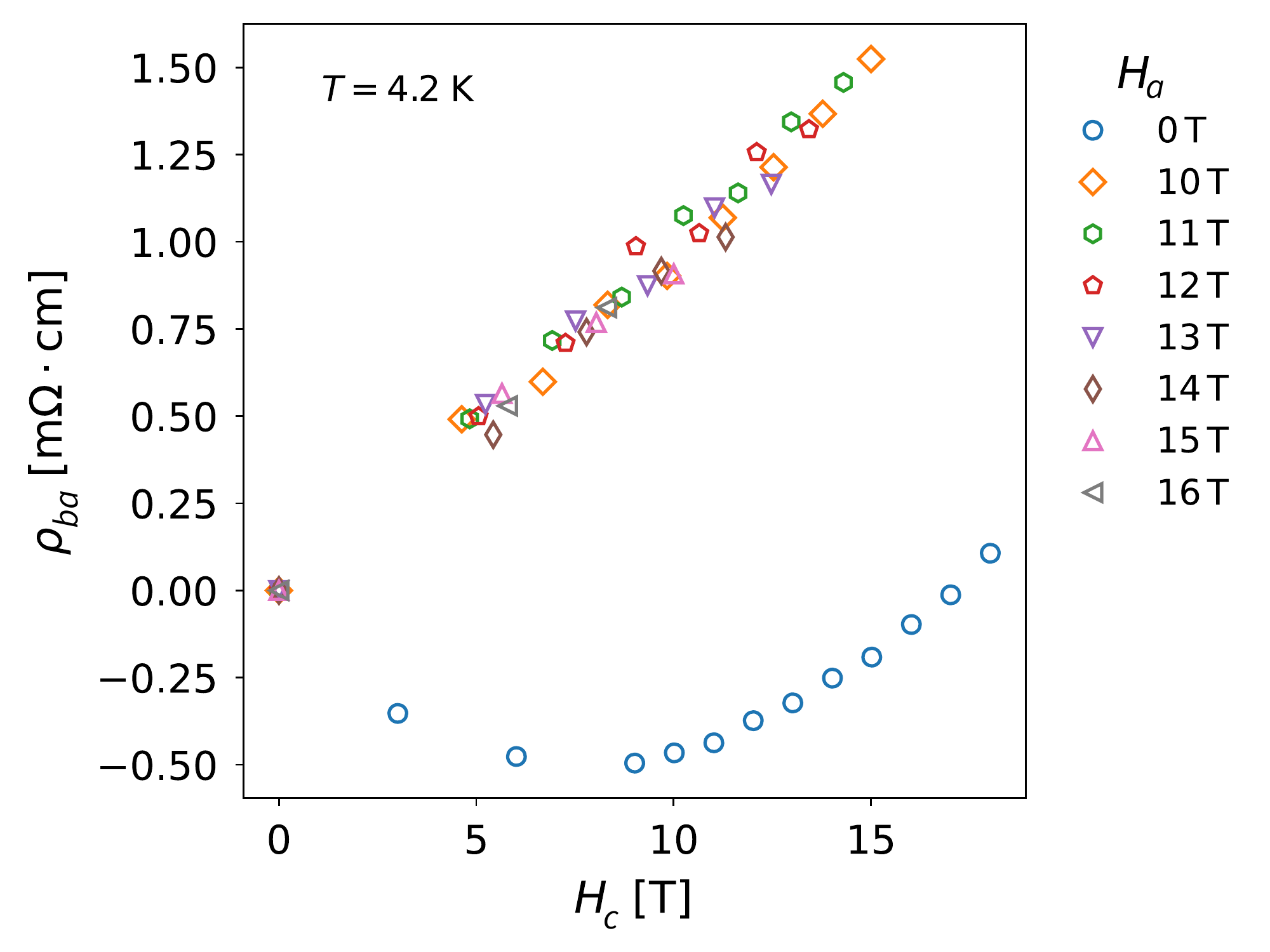}
  \caption{
    Effect of $H_a$ on the maximum value of $H_c$.
    A sharp contrast can be seen between the data at $H_a = \SI{0}{\tesla}$ and those at high $H_a$'s.
  }
  \label{fig:rhoyx-highH}
\end{figure}

Because $H_a$ and $H_c$ are not independent (Eqs.~\eqref{eq:rotate-field}), the upper limits for $(H_a, H_c)$ are smaller than the maximum magnetic field $H_{\mathrm{max}}$ that the superconducting magnet can generate.
In our case, $H_{\mathrm{max}} = \SI{18}{\tesla}$, and at $H_a = \SI{10}{\tesla}$, the $\rho_{ba}$ data are available up to $H_c = \SI{15}{\tesla}$.
If we select $H_a = \SI{11}{\tesla}$, the highest $H_c$ is reduced to approximately $\SI{14.25}{\tesla}$.
Fig.~\ref{fig:rhoyx-highH} shows that choosing higher values of $H_a$'s largely reduces the range of $H_c$ for which $\rho_{ba}$ can be measured.
On the other hand, at $H_a \geq \SI{10}{\tesla}$,  all the  $\rho_{ba}$ curves fall into a same line.
The available data are thus sufficient to reveal the Hall effect of the high $H_a$ state.

\subsection{Effects of discrete $H$ values}
\label{sec:effects-discrete-h}

\begin{figure}
  \centering
  \includegraphics[width = .9\textwidth]{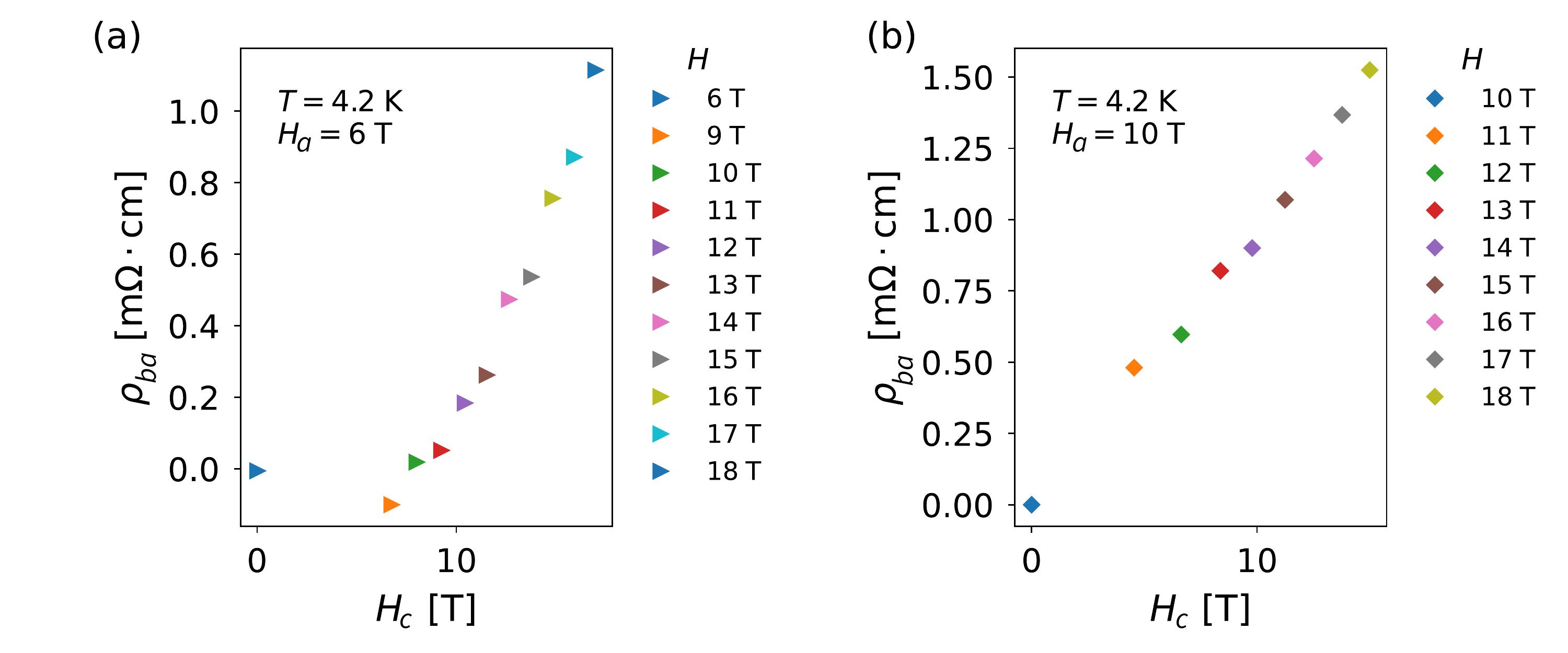}
  \caption{
    (a) $H_c$-dependence of $\rho_{ba}$ at $H_a = \SI{6}{\tesla}$ and $T = \SI{4.2}{\kelvin}$.
    Each data point is collected from the angular dependence of $\rho_{ba}(\theta)$ at a constant $H$, as shown by the legend on the right.
    (b) $H_c$-dependence of $\rho_{ba}$ at $H_a = \SI{10}{\tesla}$ and $T = \SI{4.2}{\kelvin}$.
    The number of $H$'s are different from (a).
  }
  \label{fig:rhoyx-fixedH}
\end{figure}

Since the measurements were carried out at the discrete values of $H$, gaps are unavoidable in the low $H_c$ region of the $\rho_{ba}(H_c)$ data collected at high $H_a$ [Fig.~\ref{fig:rhoyx-fixedH}].
For example, to collect $\rho_{ba}$ at $H_a = \SI{10}{\tesla}$ [Fig.~\ref{fig:rhoyx-fixedH}(b)], the minimum external field that is possible is $H = \SI{10}{\tesla}$, and we can only measure $\rho_{ba}$ at $H_c = \SI{0}{\tesla}$, i.e. at $\theta = \SI{0}{\degree}$.
At the next value, $H = \SI{11}{\tesla}$, the sample has to be tilted to $\theta = \arccos\frac{10}{11} \approx \SI{24.62}{\degree}$ to hold $H_a$ at $\SI{10}{\tesla}$, and therefore the smallest achievable $H_c$ is about $\SI{4.58}{\tesla}$.
The data between $\SI{0}{\tesla}$ and $\SI{4.58}{\tesla}$ are thus inaccessible.

Despite these unwanted gaps, the data shown in Figs.~\ref{fig:hallSM}(c) and \ref{fig:rhoyx-highH} are sufficient to demonstrate the changes from nonlinear to linear $\rho_{ba}$ curves as a function of $H_a$.
Specifically, $\rho_{ba}$'s at $H_a = \SI{0}{\tesla}, \SI{2}{\tesla}$, and $\SI{4}{\tesla}$ exhibit a similar curvature that is consistent with that observed by the published transverse Hall effect \cite{ogasawara2021}.
In the metallic state at $H_a \geq \SI{10}{\tesla}$, $\rho_{ba}$ is almost a linear line.

\subsection{Effects of tilted magnetic fields}
\label{sec:tilted-field}
In a crossed-field experiment, the total magnetic field $\vecH$ is tilted with respect to the plane made the two perpendicular vectors, namely the electric current density $\vec{J}_a$ and the Hall electric field $\vec{E}_b$.
Therefore, the measured $\rho_{ba}$, defined as,
\begin{align}
  \label{eq:rhoyx-def}
  \rho_{ba} &= \frac{E_b}{J_a}\,.
\end{align}
includes the contributions from both the transverse Hall conductivity $\sigma_{ab}$ and a second-order correction from other components of the conductivity tensor $\sigma_{ij}$.
Fortunately, for the case of \bmb{}, this correction is negligible.

In the experimental configuration shown by Fig.~\ref{fig:hallSM}(a), we have,
\begin{align}
  \label{eq:tilted-field}
  \begin{pmatrix}
    J_a \\ J_b \\ J_c
  \end{pmatrix}
  &=
    \begin{pmatrix}
      \sigma_\parallel & \sigma_{ab} & 0 \\
      -\sigma_{ab} & \sigma_{\parallel} & \sigma_{bc} \\
      0 & - \sigma_{bc} & \sigma_{cc}
    \end{pmatrix}%
                          \begin{pmatrix}
                            E_a \\ E_b \\ E_c
                          \end{pmatrix}\,.%
\end{align}
Here $\sigma_{ac} = 0$ because $\vecH$ remains in the $ac$ plane during the measurements.
Since \bmb{} adopts a tetragonal crystal structure and its magnetoresistance is isotropic in the $ab$ plane \cite{ogasawara2021}, $\sigma_{aa}$ and $\sigma_{bb}$ are equivalent, and we use $\sigma_{aa} = \sigma_{bb} \equiv \sigma_{\parallel}$.
In our settings, only $J_a \neq 0$ and $J_b = J_c \equiv 0$.
In addition, the Hall resistiviy $\rho_{ba}$ is two orders smaller than $\rho_{aa}$ \cite{ogasawara2021}, so that the approximation $\sigma_{\parallel}^2 + \sigma_{ab}^2  \approx  \sigma_{\parallel}^2$ is valid.
We then find that,
\begin{align}
  \label{eq:rhoyx-approx}
  \rho_{ba} \equiv \frac{E_{b}}{J_a} &= \frac{\sigma_{ab}}{\sigma_{\parallel}^2}%
                                   \left [1 + \frac{\sigma_{bc}^2}{\sigma_{cc}\sigma_{\parallel}} \right ]^{-1}\,.
\end{align}

In the right hand side of Eq.~\eqref{eq:rhoyx-approx}, the first factor $(\sigma_{ab} / \sigma_{\parallel}^2)$ is corresponds to the pure in-plane transverse Hall effect that arises from the $\vec{H}_a$-defined electronic state.
On the other hand, the correction in the parentheses comes from the fact that the total magnetic field $\vecH$ is tilting away from being perpendicular to both $\vec{J}$ and the $\hat{b}$-axis.
We also have $|\sigma_{ij}| = |\rho_{ij}| / ||\rho||^2$, where $||\rho||$ is the determinant of the resistivity matrix. 
The correction in the parentheses is then,
\begin{align}
  \label{eq:correction-rhoyx}
  \left [1 + \frac{\sigma_{bc}^2}{\sigma_{cc}\sigma_{\parallel}} \right ]^{-1} &=%
  \left [1 + \frac{\rho_{bc}^2}{\rho_{cc}\rho_{\parallel}} \right]^{-1}                                                                                                                         
\end{align}

From the values of the resistivity tensor of \bmb{} \cite{ogasawara2021}, we can estimate the upper limit of this correction by using the lowest values of $\rho_{aa}$ and $\rho_{cc}$ and the largest value of $\rho_{bc}$ at the same temperature ($\rho_{bc} = \rho_{ac}$  due to tetragonal symmetry).
At $T = \SI{4.2}{\kelvin}$, $\rho_{\parallel} = \rho_{aa} \approx \SI{3d-3}{\ohm\meter}$ and $\rho_{cc} \approx \SI{5d-3}{\ohm\meter}$ at $H_a = \SI{18}{\tesla}$, and $\rho_{bc} \approx \SI{2d-5}{\ohm\meter}$ at $H_a = \SI{9}{\tesla}$.
The correction $\left [ 1 + \frac{\sigma_{bc}^2}{\sigma_{cc}\sigma_{\parallel}} \right ]$ in Eq.~\eqref{eq:rhoHxHz} is then about $(1 \pm \num{2.6d-5})$.
\emph{Therefore, $\rho_{ba}$ obtained from the crossed-field measurement faithfully reflects the in-plane Hall response of the $\vec{H}_a$-defined electronic state.}

\subsection{Results and discussions}
\label{sec:Hall-result-SM}

\subsubsection{In-plane Hall effect as a function of $H_a$ and temperature}
\label{sec:plane-hall-effect}
In Fig.~\ref{fig:hallSM}\,(c) , we shows the dependencies of the current-in-plane {\itshape cip} Hall resistivity $\rho_{ba}$ versus the Hall magnetic field ($H_c$) collected at various $H_{a}$'s and temperatures.
At $T = \SI{4.2}{\kelvin}$, i.e. in the VRH regime, the $\rho_{ba}$ curve at $H_{a} = 0$ displays a non-linearity in which a negative slope at low $H_c$ bends to a positive linear
curve for $H_c \gtrsim \SI{10}{\tesla}$.
The reason for the nonlinear Hall effect is far from a competition of electronlike and holelike carriers, which is only applicable for semiclassical quasi-free electrons \cite{huynh2019,ogasawara2021}.
Instead of that, the negative Hall coefficient around $H_c = 0$ rather comes from close \emph{hopping} orbits of holelike carriers \cite{friedman1971,mott2012}.
With increasing $H_a$, the VRH-induced negative Hall effect is removed and the $\rho_{ba}(H_c)$ curves gradually flattens.
At $H_a = \SI{10}{\tesla}$, the transport regime is metallic and $\rho_{ba}(H_c)$ is almost a linear line.
At temperatures higher than the VRH regime, $\rho_{ba}(H_c)$ is linear and the slope $R_{\mathrm{H}}$ is almost unchanged slope with respect $T$ [Fig.~\ref{fig:hallSM}\,(c)].

In short, the Hall effect $\rho_{ba}$ is linear in the metallic regime at high $T$'s and/or high $H_a$'s.
The nonlinear behavior of $\rho_{ba}$ appears only in the VRH regime.
A previous study shows that the negative slope at low $H_c$ of $\rho_{ba}$ in the VRH regime is not related to a semiclassical multi-carrier-type effect \cite{ogasawara2021}.


\subsubsection{Estimations of carrier number and mobility}
\label{sec:estim-carr-numb}
We estimated the carrier number $n$ using the basic formula,
\begin{align}
  \label{eq:Hall-eq-SM}
  R_{\mathrm{H}} &= \frac{1}{en}\,,
\end{align}
where $R_{\mathrm{H}}$ is the Hall coefficient.
At high $T$'s and high $H_a$'s, i.e., outside of the VRH regime, $\rho_{ba}(H_c)$ curves are linear, and Eq.~\eqref{eq:Hall-eq-SM}  gives a reliable estimation for $n$.
On the other hand, in the VRH regime (at low $T$'s and $H_a$'s), $\rho_{ba}$ is complex and one needs more a careful approach.

As we discussed in Sec.~\ref{sec:hall-settings-general}, the negative slope in the low $H_c$ region of $\rho_{ba}$ does not correspond to an electronlike carrier type.
The presence of an electronlike carrier species was also ruled out by comparing between the results of a two-carrier-type analysis and the observations from band calculations and other physical properties \cite{ogasawara2021}.
The interpretation of $\rho_{ba}$ in the VRH regime is also difficult because there is no theoretical consensus on the behavior of a VRH transverse Hall effect as a function of the Hall magnetic field (in this case $H_c$).
We thus employed an empirical approach to study the VRH-Hall effect in \bmb{}.

The critical value $H_a$ for tuning \bmb{} to a metallic state is about $\SI{5}{\tesla}$ (see Fig.~3 in the main text and Sec. \ref{sec:scalingSM}).
At $T = \SI{4.2}{\kelvin}$ and $H_a > \SI{10}{\tesla}$, the saturating behavior of magnetoresistance of \bmb{} indicates that no other change of electronic state occurs at higher $H_a$ \cite{ogasawara2021}.
Therefore, the almost linear $\rho_{ba}$ curve found at $T = \SI{4.2}{\kelvin}$ and $H_a = \SI{10}{\tesla}$ in Fig.~\ref{fig:hallSM}(d) allows a reliable estimation of $n$ using Eq.~\eqref{eq:Hall-eq-SM}.
At the same temperature, by reducing $H_a$'s, a negative slope gradually appears as \bmb{} re-enters the VRH regime.
Because the slope of high $H_c$ region of $\rho_{ba}$ remains almost constant in this transition, we empirically used the linear tail of $\rho_{ba}$ to estimate $n$.


We thus approximated the current-in-plane (\cip{}) Hall coefficient $R_{\mathrm{H}}^{ip}$ and the carrier number $n_{ab} = 1/e R^{ip}_{\mathrm{H}}$ by estimating the slopes of the linear segments at  high-$H_c$ and found that these parameters are almost unchanged in the wide ranges of $H_a$'s and $T$'s [Fig.~\ref{fig:hallSM}\,(d)].
\emph{Therefore, the drastic change of the conductivity is mainly due to the effects of $T$ and/or $H_a$ on the mobility $\mu_{ab}$.}
Fig.~\ref{fig:hallSM}\,(e) shows the temperatures dependencies of $\mu_{ab}$ estimated for various $H_a$'s.
The solid symbols represent the values of the in-plane mobility calculated by $\mu_{ab}(H_a, T) = \sigma_{ab,0}(H_a, T)/n_{ab}e = \sigma_{ab,0} (H_a, T) R_{\mathrm{H}}^{ip}$.
Here $e$ is the elementary charge; $\sigma_{ab,0}(H_a, T)$ is the conductivity at $H_a$ and $T$, and at $H_c = 0$.

Our previous study \cite{ogasawara2021} showed that $n_{ab}$ is virtually unchanged as a function of $T$.
In addition, Fig.~\ref{fig:hallSM}(d) shows that at high temperatures, the slopes of the high $H_c$ tails in $\rho_{ba}(H_c)$ do not change with $H_a$.
These observations allow for another approximation.
We estimated the values of $\mu_{ab}$ for $H_a \geq \SI{10}{\tesla}$ by using the Hall coefficient  $R_{\mathrm{H}}^{ip} (H_a = \SI{6}{\tesla}, T = \SI{100}{\kelvin})$.
The values of $\mu_{ab}$ obtained by this approximation are shown by open symbols in Fig.~\ref{fig:hallSM}\,(d).
For $H_a = 0$, $\mu_{ab}$ displays a broad peak corresponding to the metal-to-insulator (MIT) transition occurring at $\tmin$ and then decreases to zero in entering the VRH regime.
This is consistent with the picture of an Anderson localization, in which the carrier mobility decays exponentially with decreasing temperatures \cite{mott2012}.
The application of $H_{ab}$ reverses the MIT so that $\mu_{ab}$ becomes metallic down to low $T$s.

\subsection{Current-out-of-plane Hall effect}
\label{sec:current-out-plane}

More corroborating information about the $H_{ab}$-enhanced mobility can be gathered from the current-out-of-plan ({\itshape coop}) Hall effect shown in Fig.~\ref{fig:hallSM}(d).
Here the electric current $\vec{J}$ is parallel to the $\hat{c}$-axis, and the magnetic field  $\vec{H}_a \parallel \hat{a}$ takes a dual role of being the Hall probing field and simultaneously, the tuning field that enhances the interlayer mobility $\mu_{c}$ along the $\hat{c}$-axis.
As a result, the $\rho_{bc} (H_a)$ at $T = \SI{2}{\kelvin}$ is a highly nonlinear curve which has a steep negative slope at around zero $H_a$ but quickly bends to positive.
At $H_a > \SI{15}{\tesla}$, where the MR saturates, $\rho_{bc}$ becomes linear with $H_a$.
Similar $H_a$ dependencies were observed for $\rho_{bc}$'s at elevated temperatures, and the curves always bend to linear at high $H_a$'s.

We approximated the \coop{} Hall coefficients $R_{\mathrm{H}}^{op}$ at various temperatures by the slope of the high $H_a$ linear segments.
The \coop{} carrier number $n_c$ obtained by $n_c = 1 / R^{op}_{\mathrm{H}}$ does not show any clear temperature dependence [Fig.~2(d) in the main text].
As suggested by the \cip{} Hall effect, the carrier number is unlikely to be changed by $H_a$, and this allows us to estimate the mobility at $H_a = 0$ and $\SI{16.5}{\tesla}$ by using the formula $\mu_c (H_a, T)= \sigma_{c} (H_a , T) / n_c (T) e$.
The results are shown in Fig.~2(c) in the main text.

\section{Scaling analyses}%
\label{sec:scalingSM}%
%
The scaling theory of quantum phase transition (QPT) \cite{belitz1994} assumes that a thermodynamic quantity $Q$ near the phase transition can be generalized as a homogeneous function of second order, which scales as follows;
\begin{align}
  \label{eq:q-scaling}
  Q(h,T) = b^{x_Q}Q(b^{1/\nu}h,b^zT)\,.
\end{align}
Here,  $h = H/H_{\mathrm{crit}}-1$ is the dimensionless distance measures how far the tuning parameter $H$ is from the critical point $H_{\mathrm{crit}}$.
$b$ is the scaling parameter that correspond to the scaling $r \rightarrow r/b$.
$\nu$ and $z$ are the localization length and dynamical critical exponents, respectively \cite{belitz1994}.
In our case, $Q$ and $H$ are the electrical conductivity $\sigma$ and $\Hab$.
Eq.~(\ref{eq:q-scaling}) then becomes;
\begin{subequations}
  \begin{align}
    \label{eq:sigma-scaling}
    \sigma (h,T) &= b^{-(d-2)}\mathcal{F}(b^{1/\nu}h,b^zT)\,;\\
    h &\equiv \frac{H_{ab}}{H_{ab, \mathrm{crit}}} - 1\,
  \end{align}  
\end{subequations}
with $\mathcal{F}$ is unknown scaling function.
Here, $d$ is the dimensionality of the system and $x_Q = -(d-2)$ due to the dimensionality of $\sigma$.
We focus only on the three dimensional case, so that $-(d-2) = -1$.
Eq.~(\ref{eq:sigma-scaling}) should be simplified to compare to the experimental data.%

\subsection{Zero-$T$ scaling analyses for the metallic regime}
\label{sec:zero-t-scaling}%
One way to simplify Eq.~\eqref{eq:sigma-scaling} is to choose the scaling parameter $b$ so that $b^{1/\nu}h \equiv 1$;
\begin{subequations}\label{eq:zeroT}
  \begin{align}
    \label{eq:sigma-simp2}
    \sigma (h,T)|_{b = h^{-\nu}} &= h^{\mu}\mathcal{F}\left(1,\frac{T}{h^{z\nu}}\right) \\
    \mu &= (d-1)\nu = \nu\,.
  \end{align}
 \end{subequations}

As $T \to 0$, $\mathcal{F}(1,0) \to 1$ in Eq.~(\ref{eq:sigma-simp2}).
We obtain the power law scaling for  zero-temperature conductivity:
\begin{align}
  \label{eq:zeroTpower}
  \sigma(h,T=0) \equiv \sigma_0(h) &= \mathcal{F}(1,0)h^{\nu}\,.
\end{align}

\begin{figure}
  \includegraphics[width = .45\textwidth]{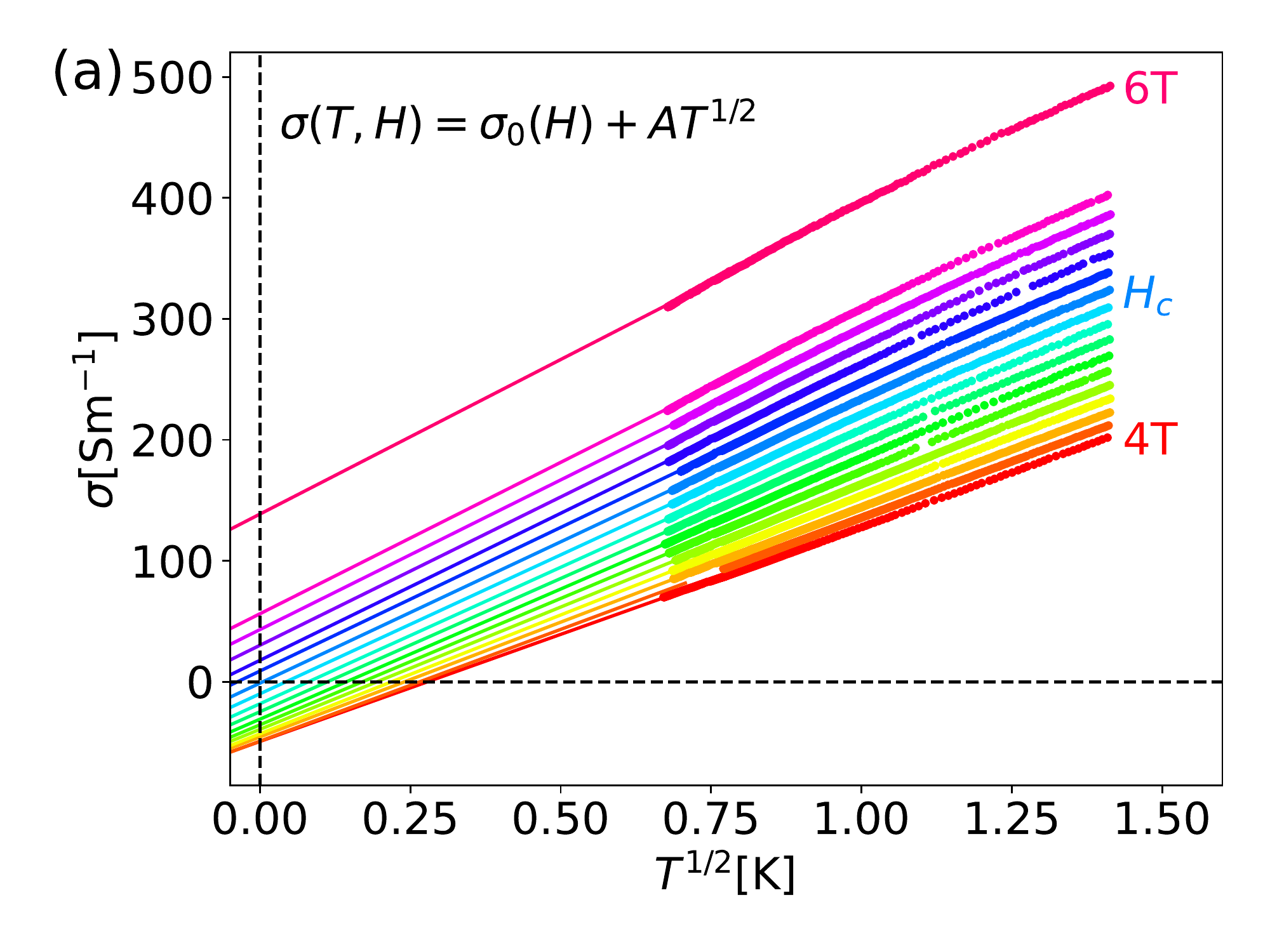}
  \includegraphics[width = .45\textwidth]{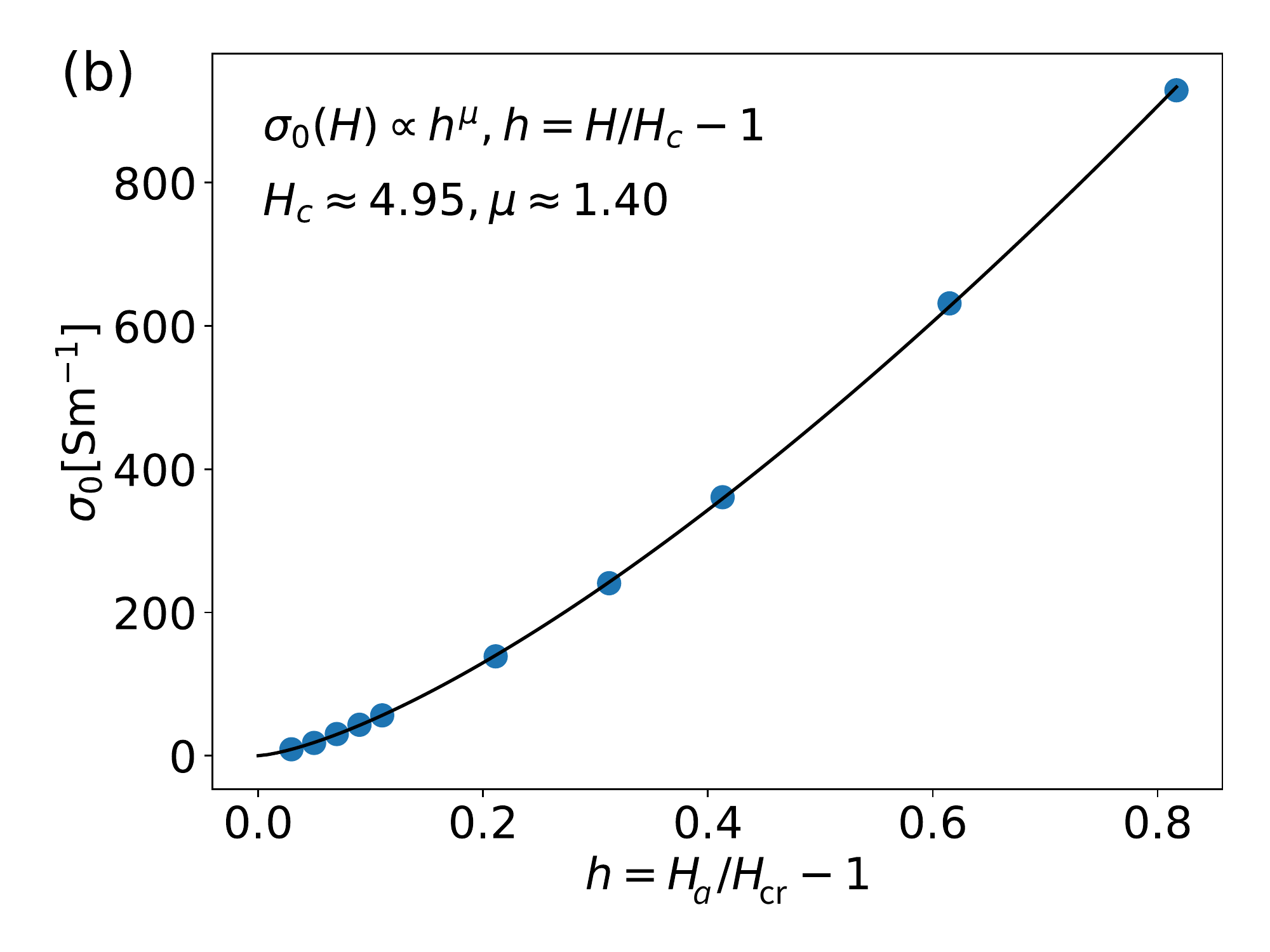}
  \caption{%
    (a) Conductivities $\sigma$'s measured at different $\Hab$'s plotted against $T^{1/2}$.
    The intercepts of the $\sigma_0 + AT^{1/2}$ fits (solid lines) defined the zero-$T$ conductivity $\sigma_0$.
    (b) Scaling of $\sigma_0(h)$ using the power law.
    The solid line is the power law fitting.}
  \label{fig:sigma0-scaling}
\end{figure}

The conductivity at zero temperature $\sigma_0$ is immeasurable, we estimated it by the following fitting; 
\begin{align}
  \label{eq:sigma-pow}
  \sigma(T,H_{ab}) = \sigma_0(H_{ab}) + AT^\alpha\,.
\end{align}
with $\sigma(T,H_{ab})$ is the experimentally measured conductivity.
The exponent $\alpha$ is chosen as to fit the experimental data, usually $1/2$ or $1/3$.
Theoretically, the value of $\alpha$ is predicted to be $1/2$ when the electron-electron interaction is taken into account.
We have tried various values of $\alpha$ to fit our data, and find that $\alpha=1/2$ minimizes the least-square residues.
Fig.~\ref{fig:sigma0-scaling}(a) show the fittings of the experimental $\sigma$'s with the model in \eqref{eq:sigma-pow} with $\alpha = 1/2$.
The $\sigma_0(h)$ data can be fitted by Eq.~\eqref{eq:sigma-pow} to obtained $\nu$ as shown in Fig.~\ref{fig:sigma0-scaling}(b).%

The procedure for the zero-$T$ scaling analyses was as follows.
\begin{enumerate*}[label = (\roman*)]
\item $\sigma_0$'s were extracted by fitting $\sigma(T)$ data measured within the window $T_{min}< T <T_{max}$ at different $H_a$'s using Eq.~\eqref{eq:sigma-pow}.
  The value of $H_a$ that corresponds to the mininum $|\sigma_0|$ gives a rough estimation of $H_{\mathrm{cr}}$.
\item fitting of the obtained $\sigma_0$ against $H_a$ using Eq.~\eqref{eq:zeroTpower} to find the critical exponent $\nu$ and the critical field $H_{\mathrm{cr}}$.
\end{enumerate*}
We fitted the data using the least-square module of the package of SciPy \cite{virtanen2020}.
The fittings in step (i) also provided the uncertainties of $\sigma_0$'s as the diagonal components of the covariance matrix that was available in the output of the fitting program.
Both $\sigma_0$'s and their accompanying uncertainties were used as the inputs for the fit in step (ii) in order to evaluate those of $\nu$ and $H_{\mathrm{cr}}$.
For S1 and S2, the $T$-fitting range was $\SI{0.5}{\K} \leq T \leq \SI{1}{\K}$.
The effects of the fitting range in $T$ and $h$ are discussed in Sec.~\ref{sec:error-analyses}.

The scaling for $\sigma_0$ cannot be used in the insulating regime $h \leq 0$ where $\sigma_0 < 0$.

\subsection{Finite $T$ scaling}
\label{sec:finite-t-scaling}
One can also modify the scaling parameter $b$ so that $b^{z}T$ becomes unity in Eq.~\eqref{eq:sigma-scaling};
\begin{subequations}
  \label{eq:finiteT-all}
  \begin{align}
    \label{eq:finiteT}
    \sigma (h,T)|_{b = T^{-1/z}} &= T^{1/z}\mathcal{F}\left(\frac{h}{T^{1/z\nu}},1\right)\,;\\
    \label{eq:betaznu}
    \beta &= z^{-1}\nu^{-1}\,.
  \end{align}
\end{subequations}
$z$ is the so-called dynamical exponent.
The finite-$T$ scaling described by Eqs.~\eqref{eq:finiteT-all} can be used to analyze the critical behaviors at both sides of $H_{\mathrm{cr}}$.

In \bmb{}, we can bring the system into the vicinity of the transition at $h = 0$ by finely tuning $\Hab$. 
The parameter $h/T^\beta$ then remains small even as $T \to 0$ and the $T$-dependence of $\mathcal{F}$ becomes negligible.
The $T$-dependence of $\sigma$ is then that of the prefactor $T^{1/z}$.

From Eq.~\eqref{eq:sigma-pow}, we have $\sigma \propto T^\alpha = T^{1/z}$, and thus,
\begin{align}
  \label{eq:alpha-z}
  \alpha &\approx z^{-1}\,.
\end{align}
Eq.~\eqref{eq:finiteT} therefore becomes,
\begin{align}
  \label{eq:finiteT-simp}
  \frac{\sigma (h,T)}{T^\alpha} & = \mathcal{F}\left(\frac{h}{T^{\beta}}\right)\,.
\end{align}
%
%
The scaling function $\mathcal{F}(h/T^\beta)$ then can be replaced by a single polynomial expansion for both insulating and metallic sides \cite{m.itoh2004};
\begin{align}
  \label{eq:finiteT-poly}
  \frac{\sigma(h, T)}{T^\alpha} & \approx  \left [%
                   a_0 + a_1\frac{h} {T^\beta} + a_2\left ( \frac{h}{T^\beta} \right )^2 + \ldots
                   \right ]\,.
\end{align}%
The analysis described in the zero-$T$ scaling analysis suggests that $\alpha\approx 1/2$ and $H_{\mathrm{cr}} \approx \SI{5.07}{\tesla}$.
In order to find $\beta$, we employed Eq.~\eqref{eq:finiteT-poly} as the fitting model for the $\sigma(h, T)$ data.
The critical exponent $\nu$ was then calculated using \eqref{eq:betaznu}.
The uncertainties of $\nu$ were evaluated directly from the output of the fit.
For S1 and S2, the $T$-fitting range was $\SI{0.5}{\K} \leq T \leq \SI{1}{\K}$.
A plot of $\sigma (h,T) / T^\alpha$ against $h/T^{\beta}$ are shown in Fig.~3(a) of the main text.
\subsection{Zero-$T$ scaling analysis for the insulating regime}
\label{sec:scal-analys-insul}
In the insulating region $h < 0$, the $T$-dependence of $\sigma$ can be modeled by a VRH law:
\begin{align}
  \label{eq:VRH-law}
  \sigma(T) &= \sigma_\infty (T)\exp\left[-\left(\frac{T_0}{T}\right)^p\right]\,,
\end{align}
where $\sigma_\infty$ is the conductivity at the $T\to\infty$ limit.
The characteristic temperature $T_0$ is \cite{mott2012},
\begin{align}
  T_0 &= \frac{24}{\pi k_BN(E_F)a^3}\,.
        \label{eq:T0}
\end{align}
where $a$ is the localization length,
$k_B$ and $N(E_F)$ are Boltzmann constant and the density of state at the Fermi level, respectively.

As $h\to 0$, the diverging localization length $a$ has the same physical meaning with that of the correlation length $\xi$, and then,
\begin{align}
  \label{eq:correlation-length}
  T_0 \propto \xi^{-3} &\propto h^{3\nu}\,.
\end{align}
The critical exponent of $T_0$ thus should be about three times of that obtained from the scaling analyses for metallic regime described in Sec.~\ref{sec:zero-t-scaling}.
As shown in the Fig.3 in the main text, the critical exponent obtained the $T_0$-scaling is consistent with the $\nu$ obtained from the scaling in the metallic regime.

\begin{figure}
  \includegraphics[width = .8\textwidth]{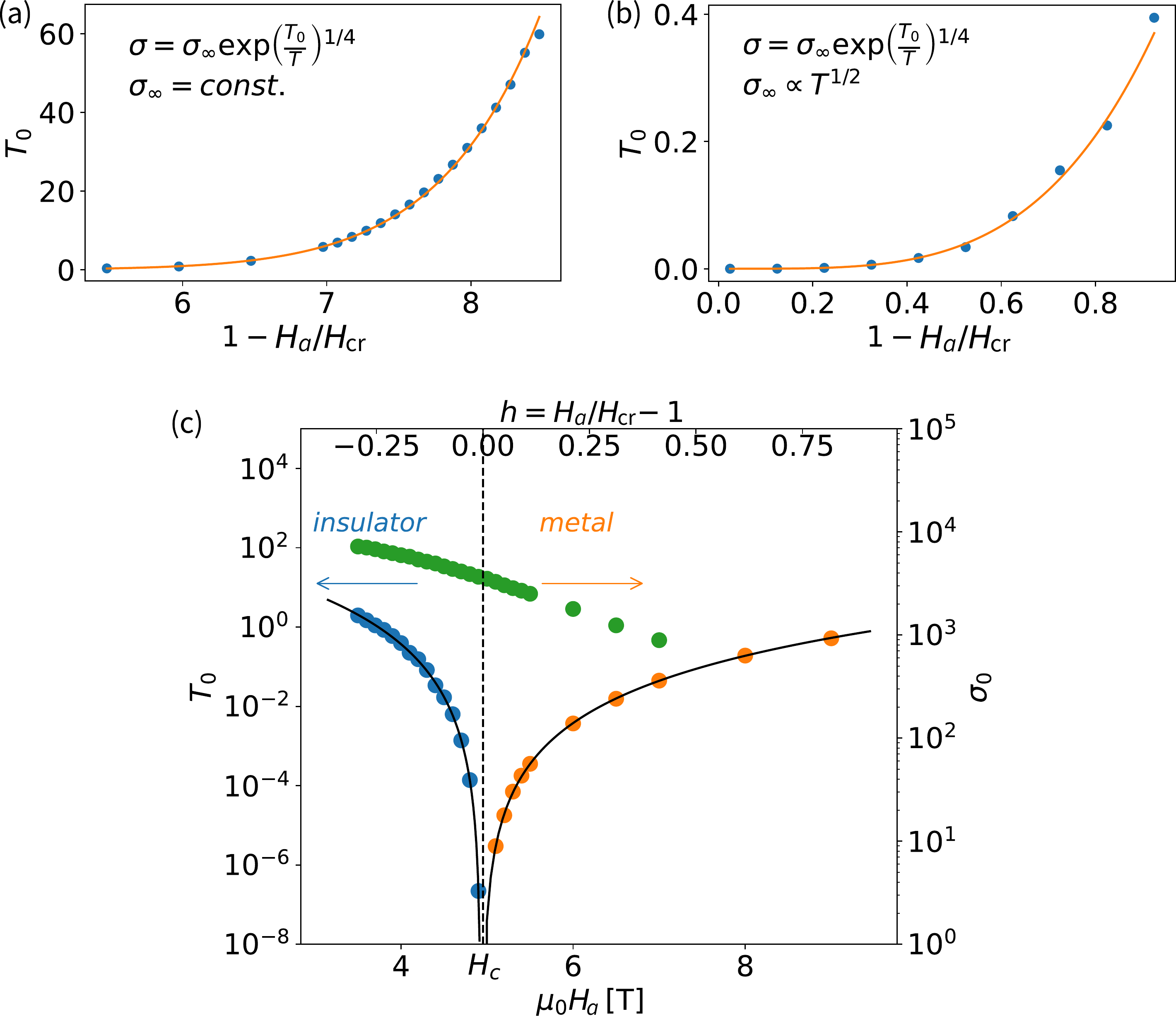}
  \caption{%
    (a) The scaling of $T_0$'s extracted from $T$-dependence of resistivity using Mott's VRH law without temperature dependent pre-factor, i.e., $\sigma_\infty = \mathrm{const}$ in Eq.~\eqref{eq:VRH-law}.
    (b) The scaling of $T_0$'s obtained from the VRH law with $\sigma_\infty \propto T^{1/2}$.
    (c) A comparison between scaling different $T_0$'s and $\sigma_0$ in the metallic regime.
    The green circles represent the for $\sigma_\infty = \mathrm{const}$.
    The blue circles represent the $T_0$ obtained from the modified VRH law with $\sigma_\infty \propto T^{1/2}$.
  }
  \label{fig:T0-scaling-all}
\end{figure}

We note that in order to extract $T_0$ from the $T$-dependencies of the conductivities in the insulating regime in the vicinity of the critical point, we employed a modified VRH law;
\begin{align}
  \label{eq:mod-VRH}
  \sigma(T) &\propto T^{q}\exp\left[-\left(\frac{T_0}{T}\right)^{p}\right]\,,
\end{align}
where $p = \frac{1}{4}$ and $q = \frac{1}{2}$.
Eq.~(\ref{eq:mod-VRH}) is thus a 3-dimensional VRH law with the prefactor $\sigma_\infty$ that varies as $T^{1/2}$ (Fig.~\ref{fig:T0-scaling-all}(b)).

In general, the $T$-dependence of the prefactor $\sigma_\infty$ is often negligible because the exponential factor converges to zero much faster as $T$ decreases.
However, the localization length $a$ diverges at the metal insulator transition, so that $T_0 \to 0$ following Eq.~\eqref{eq:T0}.
Therefore, $\sigma_\infty \propto T^q$ becomes important.
Furthermore, in the vicinity of the critical point, Eq.~\eqref{eq:mod-VRH} should approach $\sigma \propto T^q$ of the metallic regime.
In our article, the exponents $q = 1/4$ and $p=1/2$ gave the best fits to experimental data.
We also made a trial analysis using the VRH law with $\sigma_\infty = \mathrm{const}$.
Although $\sigma(T)$'s could be fitted, the scaling of the obtained $T_0$'s yielded a large critical field and is inconsistent with the analyses for the metallic regime (see Figs.~\ref{fig:T0-scaling-all}(a) and (c)).

In the scaling procedure for the insulating regime at $T = \SI{0}{\kelvin}$, at first Eq.~\eqref{eq:mod-VRH} were used as the fit model to extract $T_0$'s from the $\sigma(T)$ measured at different $H_a$'s.
For S1 and S2, the $T$-fitting range was $\SI{0.5}{\K} \leq T \leq \SI{1}{\K}$.
The obtained $T_0$'s and their uncertainties were fit by Eq.~\eqref{eq:correlation-length} to find $H_{\mathrm{cr}}$ and $\nu$ and their uncertainties.

\subsection{Uncertainties of the results}
\label{sec:error-analyses}
\subsubsection{Sample dependence}
\label{sec:sample-dependence}
We measured $T$-dependence of the conductivity of six single crystalline samples.
The results of the scaling analyses for these samples are summarized in Table.~\ref{tab:exponents}.
Despite the numbering, S1 and S2 were measured after the test measurements and analyses had been carried out for S3-6.
The lowest $T$'s in the measurements for S1 and S2 were $\SI{0.5}{\K}$, an that for S3-6 were $\SI{2}{\K}$.
The lower $T$-range put S1 and S2 closer to the QCP at $T = \SI{0}{\K}$, therefore, the scaling analyses for them produce more reliable results.
The results of S1 are shown in the main text.

The value of $H_{\mathrm{cr}}$ slightly varies  between S1 and S2, perhaps due to a small difference in the carrier number.
We note that \bmb{} is at the vicinity of a QCP, where a tiny change in the carrier number may cause a substantial variation of $H_{\mathrm{crit}}$.
This sample dependence does not affect the values of the critical exponent $\nu$.
Both zero- and finite $T$ scaling analyses in the metallic regime of both samples yield consistent values of $\nu$.
The large errors in the results of the $T_0$ scaling analyses are discussed below.

We also note that the value of $\nu$ depends on the specific tuning parameter inducing the transition.
Hence, the criticality of the $h$-induced transition is generally different from that controlled by another physical parameter, such as carrier number $n$, strain, or pressure.
On the other hand, when the system is in the vicinity of the phase boundary in the plane made by two tuning parameters, e.g. $h$ and $n$, a power law relation $n \propto h^\delta$ can exist.
The scaling of either $h$ or $n$ will then give a similar critical exponent \cite{watanabe1999}.
For \bmb{}, controlling carrier density is difficult because the potassium doped samples decompose quickly in the air.

\begin{table}
  \caption{\label{tab:exponents}
    Summary of the critical exponents and errors estimated from three scaling analyses: $T_0 \propto h^{3\nu}$ for the insulating regime,
    zero-$T$ conductivity $\sigma_0 \propto h^\mu$,
    and finite-$T$ $\sigma(h,T) = T^\alpha\mathcal{F}(h/T^\beta)$ in the metallic regime.
    Here $\mu \equiv \nu$ and $\alpha\beta^{-1} = \nu$.
    The $T$-ranges of fitting are $\SI{0.5}{\K} \leq T \leq \SI{1}{\K}$ for S1 and S2, and $\SI{2}{\K} \leq T \leq \SI{3}{\K}$ for S3-6.
  }
  \begin{tabularx}{\textwidth}{@{}Y|YY|YY|Y@{}} \hline
    \multicolumn{1}{c|}{}&\multicolumn{2}{c|}{$T_0(h)$} &\multicolumn{2}{c|}{$\sigma_0(h)$}&\multicolumn{1}{c}{$\sigma(h,T\neq0)$} \\
    No. &     $\nu$  &      $\mu_0H_{\mathrm{cr}}[\si{T}]$ &        $\mu \equiv \nu$ &  $\mu_0H_{\mathrm{cr}}[\si{T}]$ &     $\nu \equiv \alpha\beta^{-1}$   \\ \hline \hline
    S1 &  1.40 $\pm$   0.07 & 5.09 & 1.39  $\pm$ 0.01 & 5.05 &    1.415 $\pm$   0.002 \\
    S2 &  1.42 $\pm$   0.07 & 4.36 & 1.38 $\pm$  0.01 & 4.32 &    1.408 $\pm$   0.001 \\ \hline
    S3 &  1.5  $\pm$   0.3  & 4.77 & 1.43 $\pm$  0.01 & 4.68 &    1.35  $\pm$   0.02 \\
    S4 &  1.5  $\pm$   0.5  & 4.8  & 1.57 $\pm$  0.01 & 4.58 &    1.50  $\pm$   0.04 \\
    S5 &  1.2  $\pm$   0.5  & 4.54 & 1.43 $\pm$  0.02 & 3.76 &    1.25 $\pm$    0.02 \\
    S6 &  1.6  $\pm$   0.4  & 4.31 & 1.41 $\pm$  0.01 & 4.19 &    1.30 $\pm$    0.02 \\ \hline
  \end{tabularx}
\end{table}

\subsubsection{Effect of positive magnetoresistane component in the VRH regime}
\label{sec:effect-posit-magn}

As shown in Figs.\,~\ref{fig:hallSM}(b) and (g), in the VRH regime, \bmb{} shows a complex magnetoresistance (MR) that consists of competing positive and negative components \cite{huynh2019,ogasawara2021,jansa2021}.
The positive MR saturates at $\Hab \gtrsim \SI{2}{\tesla}$, from which the negative MR dominates until itself enters a saturation at $\Hab > \SI{10}{\tesla}$.
As $T$ increases, the positive MR quickly decays, and only the negative MR exists out of the VRH regime.

The positive MR has a negligible effect for the $\sigma_0$- and finite-$T$ scaling analyses.
The $\sigma_0$ scaling is used in the metallic regime, in which the negative MR dominates the conduction, and the finite-$T$ scaling focuses on relatively small windows around $H_{\mathrm{cr}} \approx \SI{5}{\tesla}$.
On the other hand, for $T_0$-scaling analyses, the positive MR introduces rather larger uncertainties into the results.
Its strong $T$-dependence affects the evaluation of $T_0$'s via Eq.~\eqref{eq:mod-VRH}.
As a results, the error bars of the scaling results are sensitive to the temperature fitting range (Fig.~\ref{fig:fitting-range}(a)).
The positive MR also forced us to use only the $T_0(h)$ points near $H_{\mathrm{cr}}$ to reduce the effect of positive MR.

\subsubsection{Effects of $T$ and $H_a$ fitting ranges}
\label{sec:effect-temp-fitt}
\begin{figure}
  \includegraphics[width = .9\textwidth]{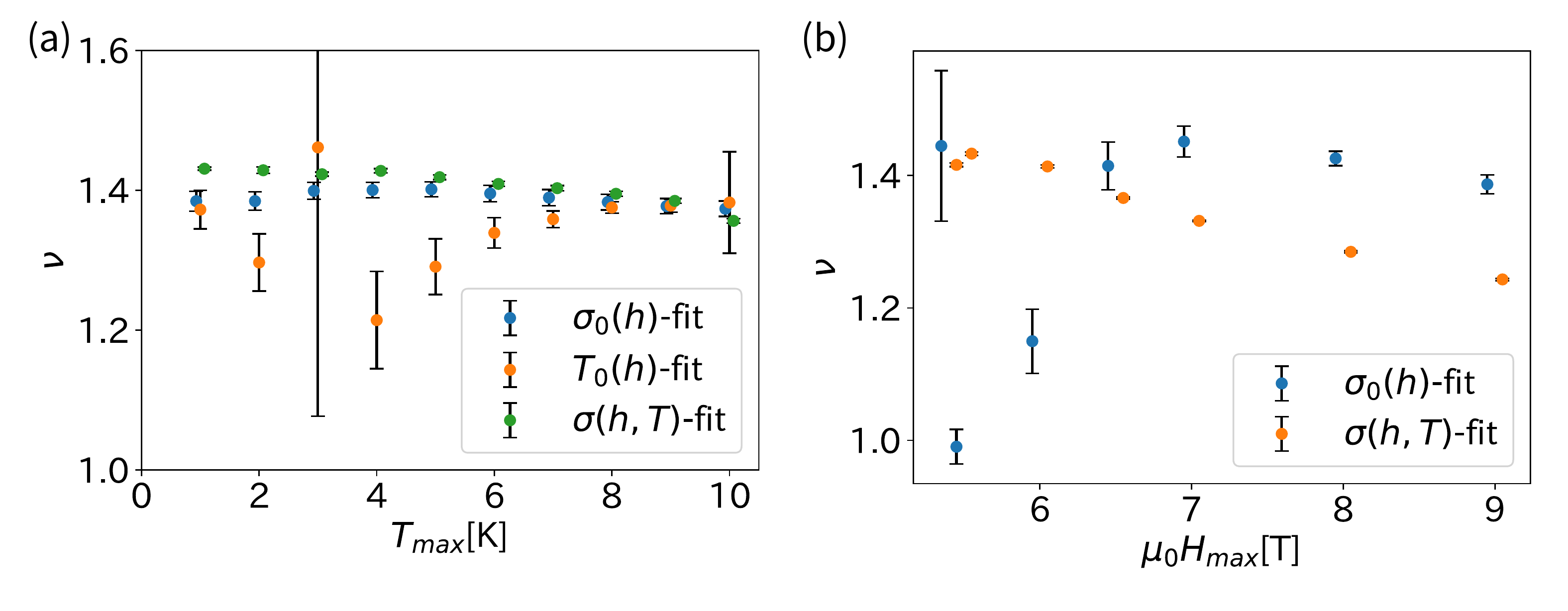}
  \caption{The dependencies of $\nu$ and its errors on the fitting windows in (a) $T$ and (b) $H_a$.}
  \label{fig:fitting-range}
\end{figure}

Fig.~\ref{fig:fitting-range}(a) shows effects of the fitting ranges in $T$ on the results of the scaling analyses of S1.
The lower limit of the fitting window is $T_{min} = \SI{0.5}{\kelvin}$, and upper limit $T_{max}$ is varied from $\SI{1}{\kelvin}$ to $\SI{10}{\kelvin}$.
For the $T_0$ scaling, we use the data in the $H_a$-window $\SI{3}{\tesla} \leq H_a \leq \SI{4.9}{\tesla}$.
The effects of the strong $T$-dependence of the positive MR can be seen clearly in the $T_0$ scaling.
The values of $\nu$ obtained from the $T_0$ scaling fluctuate with increasing $T_{max}$ and have large error bars.
On the other hand, for the $\sigma_0(h)$ and $\sigma(h,T)$ scaling analyses, the value $\nu \approx 1.4$ can be consistently achieved even in the extended $T$-range of $\SI{0.5}{\K}\leq T \leq \SI{10}{\K}$.
The error bars for these analyses are almost unaffected by changing $T$-fitting range.
In Fig.~\ref{fig:fitting-range}(a), the $H_a$-windows are $\SI{5.07}{\tesla} \leq H_a \leq \SI{9}{\tesla}$ and  $\SI{4.5}{\tesla} \leq H_a \leq \SI{6}{\tesla}$ for the $\sigma_0$ and the finite-$T$ analyses, respectively.

Fig.~\ref{fig:fitting-range}(b) shows effects of the fitting ranges in $H_a$ on the results of the  the $\sigma_0$ and finite-$T$ scaling analyses performed using the $T$-window of $\SI{0.5}{\K} < T < \SI{1}{\K}$ for S1.
For the $\sigma_0$ scaling using Eq.~\eqref{eq:zeroTpower}, we fix the lower limit $H_{min}$ at $H_{\mathrm{cr}} \approx \SI{5.07}{\tesla}$ and vary  the upper limit $H_{max}$.
The blue points in Fig.~\ref{fig:fitting-range}(b) show that $\nu$ fluctuates in the region of small $H_{max}$, but then converges to the value $\nu \approx 1.4$.
The error bars resulted from this analysis also shrink as $H_{max}$ increases.
The effects of $H_{max}$ in this analysis is rather trivial and simply due to the small number of data points available for the fittings carried out at small $H_{max}$'s.

We investigate effect of $H_a$-window to the result of the finite-$T$ scaling (Eq.~\eqref{eq:finiteT-simp}) by fixing $H_{min} = \SI{4}{\tesla}$ and varying $H_{max}$.
The orange points in Fig.~\ref{fig:fitting-range}(b) show that the value of $\nu$ obtained from the finite-$T$ scaling decreases with widening the $H_a$-window.
This effect can be explained as follows.
The finite-$T$ analyses employ the polynomial expansion shown in Eq.~\eqref{eq:finiteT-poly}, the validity of which relies on condition that the variable $h/T^\beta$ has to be sufficiently small.
By increasing $H_{max}$, the $\sigma(T)$ data with large $h/T^\beta$ are included into the fitting.
For example, at $H_a = \SI{9}{\tesla}$, given that $\beta = 0.355$ and $H_{\mathrm{cr}} \approx \SI{5.07}{\tesla}$, $h/T^\beta$ varies in the window $0.77 \leq h/T^\beta \leq 0.99$.
The data at such large values of $h/T^\beta$ can make the expansion becomes unsuitable.
In our analysis shown in the main text, the $H_a$-window is $\SI{4.5}{\tesla} \leq H_a \leq \SI{6}{\tesla}$, and hence $-0.144
\leq h/T^\beta \leq 0.235$.

\bibliographystyle{apsrev4-1}
\bibliography{BaMn2Pn2_related.bib}